%% file: main.tex
\documentclass[aps,prl,twocolumn,nofootinbib,groupedaddress,superscriptaddress
,longbibliography]{revtex4-2}
\include{header}

\usepackage{amsmath,amssymb}
\usepackage{multirow}
\usepackage{bm}
\usepackage{mathtools}
\usepackage{dsfont}
\usepackage{amsfonts}
\usepackage[normalem]{ulem}
\usepackage{url}
\usepackage{wasysym}
\usepackage{bbm}

\usepackage[colorlinks=true,linkcolor=magenta,citecolor=magenta,hypertexnames=false]{hyperref}

\def\ba#1\ea{\begin{align}#1\end{align}}
\def\bg#1\eg{\begin{gather}#1\end{gather}}
\def\bpm{\begin{pmatrix}}
\def\epm{\end{pmatrix}}

\newcommand{\magenta}[1]{\textcolor{magenta}{#1}}

\newcommand{\ourtitle}{
Fractionalization Signatures in the Dynamics of Quantum Spin Liquids}

\allowdisplaybreaks

\begin{document}
\title{\textbf{\ourtitle}}

\author{Kang Wang}
\altaffiliation{These authors contributed equally to this work.}
\affiliation{Beijing National Laboratory for Condensed Matter Physics and Institute of Physics,
Chinese Academy of Sciences, Beijing 100190, China}
\affiliation{School of Physical Sciences, University of Chinese Academy of Sciences, Beijing 100049, China}

\author{Shi Feng}
\altaffiliation{These authors contributed equally to this work.}
\affiliation{Department of Physics, The Ohio State University, Columbus, Ohio 43210, USA}

\author{Penghao Zhu}
\affiliation{Department of Physics, The Ohio State University, Columbus, Ohio 43210, USA}

\author{Runze Chi}
\affiliation{Beijing National Laboratory for Condensed Matter Physics and Institute of Physics,
Chinese Academy of Sciences, Beijing 100190, China}
\affiliation{School of Physical Sciences, University of Chinese Academy of Sciences, Beijing 100049, China}

\author{Hai-Jun Liao}
\affiliation{Beijing National Laboratory for Condensed Matter Physics and Institute of Physics, Chinese Academy of Sciences, Beijing 100190, China}
\affiliation{Songshan Lake Materials Laboratory, Dongguan, Guangdong 523808, China}

\author{Nandini Trivedi}
\email{trivedi.15@osu.edu}
\affiliation{Department of Physics, The Ohio State University, Columbus, Ohio 43210, USA}

\author{Tao Xiang}
\email{txiang@iphy.ac.cn}
\affiliation{Beijing National Laboratory for Condensed Matter Physics and Institute of Physics, Chinese Academy of Sciences, Beijing 100190, China}
\affiliation{School of Physical Sciences, University of Chinese Academy of Sciences, Beijing 100049, China}
\affiliation{Beijing Academy of Quantum Information Sciences, Beijing, China}


\begin{abstract}
We investigate the signatures of fractionalization in quantum spin liquids by studying different phases of the Kitaev honeycomb model in the presence of an out-of-plane magnetic field through which the model becomes non-integrable. Using the infinite projected entangled pair states (iPEPS) ansatz, along with analytical calculations and exact diagonalization, we calculate dynamical signatures of fractionalized particles through spin-spin and dimer-dimer correlations. Our analysis demonstrates the ability of these correlations to discern distinct fractionalized quantum sectors, namely Majorana fermions and the emergent $Z_2$ fluxes, in both the chiral spin liquid (CSL) phase under weak field and the emergent intermediate gapless phase (IGP) under moderate field. Importantly, our calculation reveals the nature of IGP observed at moderate fields, a region of ongoing debate, indicating that this phase is a Majorana metal induced by strong flux fluctuations. 
\end{abstract}

\maketitle

\let\oldaddcontentsline\addcontentsline
\renewcommand{\addcontentsline}[3]{}

Fractionalization can arise from strong frustration between localized spins, representing a hallmark of quantum emergent phenomena \cite{JVoit1995, Sondhi2012,knolle2015dynamics, Bloch2020}. A classic illustration of this is found in two-dimensional Mott insulators where spins become frustrated due to the spin-orbit coupling. In this scenario, each spin experiences conflicting exchange interactions from its neighboring spins \cite{Trebst2022}, preventing the formation of conventional spontaneous symmetry-breaking order. The ground state thus exhibits a quantum spin liquid (QSL) phase characterized by fractionalized degrees of freedom, with other intriguing properties such as braiding statistics and long-range entanglement~\cite{Wen1990,wen2002quantum,kitaev2006anyons,Tao07,Wen2010,savary2016quantum,ZhouRMP,Knolle2018,KnolleARCMP2019,KHATUA20231}. 
Despite significant efforts dedicated to finding candidate QSL materials such as $\alpha\text{-}$RuCl$_3$ \cite{plumb2014alpha,sandilands2015scattering,banerjee2017neutron,ywq2017,do2017majorana,Kasahara2018,wen2018,Czajka2023,yang2023point}, the quest to explicate fractionalization and its observable consequences, especially when the system is outside the scope of exact solution of integrable models \cite{kitaev2006anyons,Tao07}, remains one of the most formidable challenges in both theory and experiment.

In this Letter, we propose experimentally testable signatures of fractionalization through dynamical higher-order spin correlations. Specifically, we focus on the Kitaev honeycomb model \cite{kitaev2006anyons} in a magnetic field applied out of the plane. 
In experiments on QSL candidate materials, often an external magnetic field is required to suppress magnetic order at low temperatures due to non-Kitaev exchange interactions 
\cite{Kasahara2018,Takagi19,czajka2021oscillations,Zhao2022CPL,Czajka2023,Zhang23,Matsuda23}.
We implement the infinite projected entangled pair states (iPEPS) ansatz~\cite{verstraete2004renormalization,vanderstraeten2015excitations,vanderstraeten2019simulating,ponsioen2020excitations,ponsioen2022automatic,chi2022spin} to investigate both single-spin flip or spin-spin dynamical correlation function $S_1(\mathbf{k},\omega)$ and two-spin flips or dimer-dimer correlation function $S_2(\mathbf{k},\omega)$. The iPEPS approach provides unprecedented energy and momentum resolution, significantly surpassing conventional methods like exact diagonalization (ED) and density matrix renormalization group.
Our main results are the sharp signatures of fractionalization that form our main predictions for inelastic neutron scattering (INS) and resonant inelastic x-ray scattering (RIXS) experiments~\cite{Kumar2018,Schlappa2018}. We demonstrate that dynamical dimer-dimer correlations $S_2(\mathbf{k},\omega)$ show definitive signatures of fractionalization compared to $S_1(\mathbf{k},\omega)$.  In addition, we provide solid evidence for the gapless nature of a controversial intermediate spin liquid phase.

In the chiral spin liquid (CSL) phase we find (i) whereas single spin flip spectra mix features of Majorana fermions and fluxes and show fuzzy features~\cite{Knolle2014}, the two spin-flip spectra separate out the fractionalized quantum sector of Majorana fermions from the emergent $Z_2$ fluxes. (ii) Certain components of the two spin-flip spectra show definitive dispersive modes, attributable to Majorana fermions. Our iPEPS spectra agree with analytical calculations within perturbation theory. 
In the heavily debated intermediate gapless phase (IGP) under moderate field \cite{Zhu_PRB_2018,Gohlke_PRB_2018,LiangPRB2018,Nasu_PRB_2018,Jiang_arXiv_2018,gohlke2018dynamical,liu2018dirac,David2019,Kaib2019PRB,hickey2019emergence,Patel12199,Teng20,PradhanPRB2020,zhang2023machine,ZhangNatComm2022,Jiang2020,Han21,Baskaran2023,feng2023dim,holdhusen2023}, we find (i) both $S_1(\mathbf{k},\omega)$ and $S_2(\mathbf{k},\omega)$ obtained by iPEPS confirm its gapless spectrum down to the low-energy scale which is lower than putative gaps obtained by previous parton mean field theories \cite{Jiang2020,ZhangNatComm2022}. (ii) Signatures of fractionalization are seen in $S_1(\mathbf{k},\omega)$ which yields a very broad continuum signal. (iii) $S_2(\mathbf{k},\omega)$, remarkably, shows considerably sharper features at low energies despite fractionalization. (iv) Supported by data from iPEPS and ED, we present arguments that this IGP is a gapless Majorana metal phase induced by fluctuations of the $Z_2$ gauge field, which exhibits a log divergence in the Majorana density of states at low energy.

\begin{figure}
    \centering
    \includegraphics[width=\linewidth]{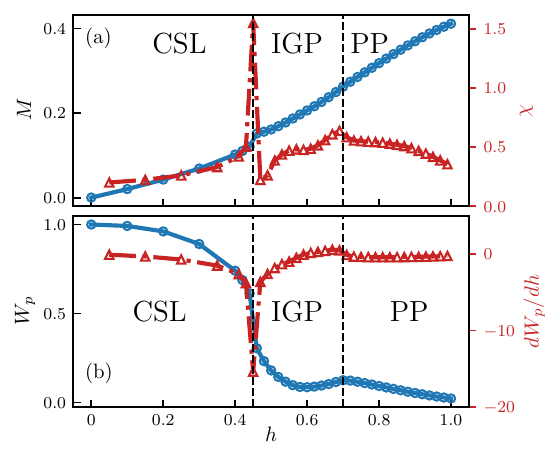}
\caption{Phase diagram measured by (a) ground-state magnetization $M$, (b) expectation of flux $W_p$, and their derivatives with respect to the field $h$ perpendicular to the plane. Kinks at $h_{c1} \simeq 0.45$ and $h_{c2} \simeq 0.70$ mark the transition from the chiral spin liquid (CSL) to the intermediate gapless phase (IGP), and from IGP to the polarized phase (PP). All data are obtained by iPEPS with bond dimension $D = 5$. }
    \label{fig:diagram}
\end{figure}

\magenta{\it Phase diagram by iPEPS .| } 
The Kitaev model under a field along [111] is depicted by the Hamiltonian: 
\begin{align}
    H = \sum_{\langle ij\rangle, \alpha}{K_\alpha \sigma^\alpha_i \sigma^\alpha_j} - h\sum_{i,\alpha}\sigma_i^\alpha ,~\alpha\in \{x,y,z\}
    \label{eq_ham}
\end{align}
We focus on the isotropic antiferromagnetic compass exchange and set $K_\alpha = 1$. We employ iPEPS as the variational ansatz for the eigenstates of Eq.~(\ref{eq_ham}) and obtain the phase diagram.
The ground state is obtained by minimizing the energy on the effective square Bravais lattice through automatic differentiation techniques \cite{liao2019differentiable}.
Excited states are obtained using the variational ansatz in the momentum representation $
\ket{\Psi_\mathbf{k}}=\sum_{\mathbf{r}}e^{i\mathbf{k}\mathbf{r}}\ket{\Psi_\mathbf{r}}$, 
where $\ket{\Psi_\mathbf{r}}$ is the state with site $\mathbf{r}$ being excited
(see details of iPEPS in Supplemental Materials \cite{re:supp}). We fix the bond dimension to $D=5$ and the boundary bond dimension to $\lambda = 100$. The phase diagram is illustrated in Fig.~\ref{fig:diagram}. We utilize the total magnetization $M$ and the flux operator $W_p=\sigma_1^x\sigma_2^y\sigma_3^z\sigma_4^x\sigma_5^y\sigma_6^z$ to characterize the phase transition. The magnetization can be optimized down to $10^{-4}$ and $W_p$ up to $0.99$ for the pure Kitaev model, in agreement with known results \cite{iregui2014probing,lee2019gapless,lukin2023variational,liao2019differentiable,zhang2023differentiable}.
The magnetic susceptibility $\chi$ and the derivative of $W_p$ peak at $h_{c1}\simeq 0.45$ and $h_{c2}\simeq 0.70$, indicating two phase transitions, one between CSL and IGP, and the other between IGP and the polarized phase (PP)~\cite{footnote}  Specifically,
upon applying a small magnetic field, the Majorana fermions develop a gap and acquire a non-zero Chern number, and meanwhile, the gauge fluctuation reduces $W_p$. When $h_{c1} < h < h_{c2}$, the perturbative picture breaks down due to the strong gauge fluctuations, and $W_p$ shows a sharp decrease, marking the system's entry into the IGP. Eventually, when $h>h_{c2}$, the system becomes polarized.  The critical points at $h_{c1}$ and $h_{
c2}$ are in close agreement with those reported in previous studies based on finite-size numerics \cite{gohlke2018dynamical,hickey2019emergence,David2019,Patel12199}. 

\magenta{\it Dynamical Spectra in CSL .| }
In this section we describe and analyze iPEPS results for the single- and two-spin flip structure factors in the low-field case. The single-spin flip structure factor is defined by
\begin{equation}
    S_1^\alpha(\mathbf{k},\omega) = \sum_{m \neq 0} \mel{0}{\sigma_\mathbf{k}^\alpha }{m}\mel{m}{\sigma_{-\mathbf{k}}^\alpha}{0} \delta(\omega - E_m + E_0),
\end{equation}
where $E_m$ stands for the energy of the $m$-th excited state $\ket{m}$; and the two-spin flip structure factor by
\begin{equation}
    S_2^\alpha(\mathbf{k}, \omega) = \sum_{m \neq 0} \mel{0}{\mathcal{D}_\mathbf{k}^\alpha }{m}\mel{m}{\mathcal{D}_{-\mathbf{k}}^\alpha}{0} \delta(\omega - E_m + E_0),
\end{equation}
where we introduced the notation $\mathcal{D}_j^\alpha \equiv \sigma_{j}^\alpha \sigma_{j+z}^\alpha$ for the two-spin dimer operators, and $\mathcal{D}_{\mathbf{k}}^{\alpha}$ is the $\mathcal{D}_j^\alpha$ transformed into momentum space.
These are useful probes for studying Kitaev materials via INS and RIXS experiments, with the latter able to capture high-order processes like two-spin-flip, i.e., four-spinon excitations \cite{Kumar2018,Schlappa2018}. 
One of our important results computed by the state-of-the-art iPEPS algorithm is that in the CSL phase $S_2(\mathbf{k},\omega)$ distinctly isolates Majorana fermion features that can be separated from the flux excitations. Our results are obtained by iPEPS, with insights from analytical calculations.

\begin{figure}
    \centering
    \includegraphics[width=\linewidth]{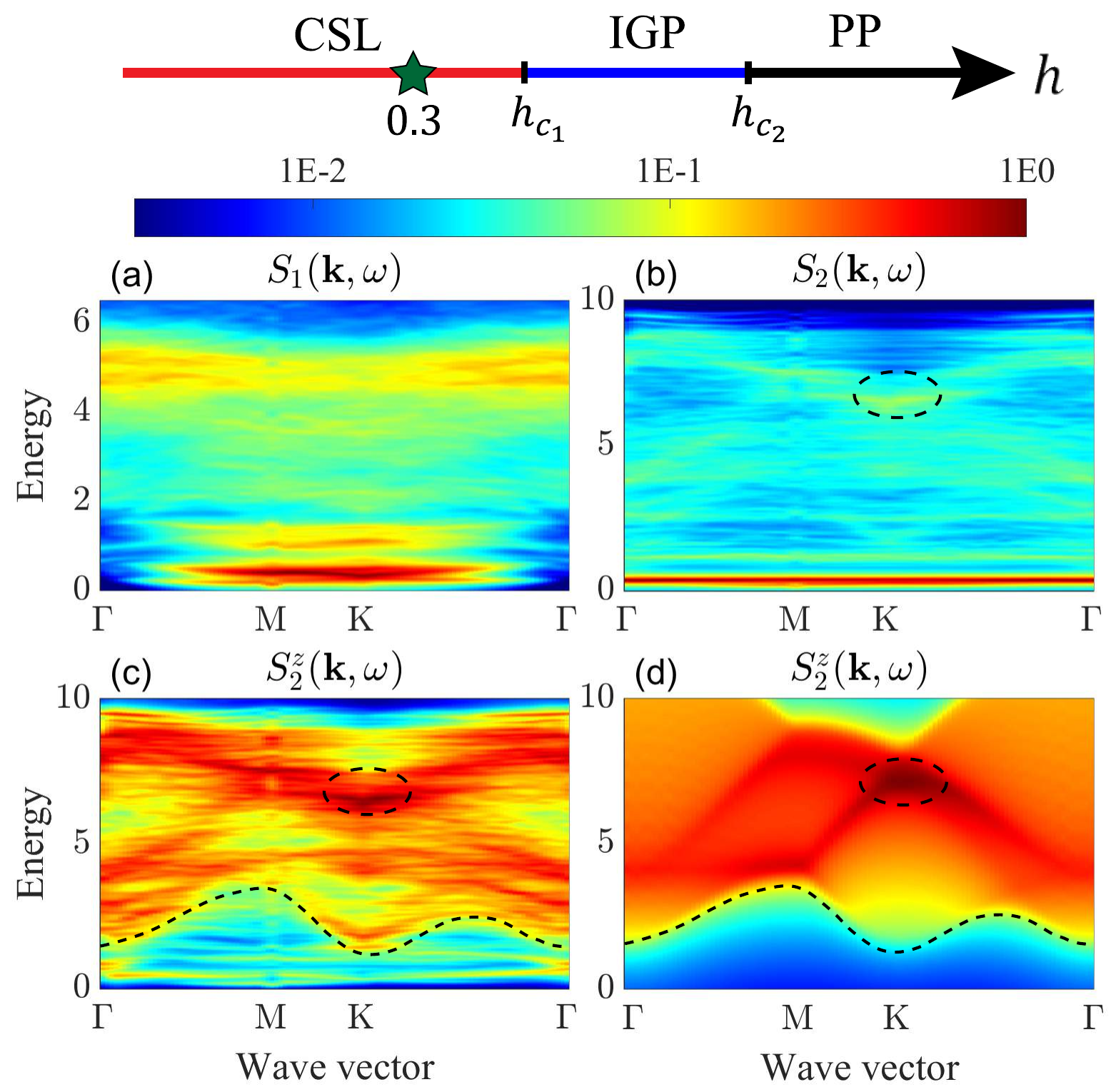 }
    \caption{Dynamical structure factors of the antiferromagnetic Kitaev model in the weak-field-induced CSL phase. (a) Total spin-spin spectrum $S_1(\mathbf{k},\omega) = \sum_{\alpha} S_1^{\alpha} (\mathbf{k},\omega)$, (b) total dimer-dimer spectrum $S_2(\mathbf{k},\omega) = \sum_{\alpha} S_2^{\alpha} (\mathbf{k},\omega)$, and (c) the dimer-dimer spectrum along $z$ axis $S_2^{z}(\mathbf{k},\omega)$. Spectra are obtained using iPEPS with the bond dimension $D=5$ and a Lorentzian broadening factor $\eta = 0.05$. (d) Analytical result obtained by the leading order time-reversal-breaking perturbation $g=0.08$, as defined in $Q_\mathbf{k}$ after Eq.~\eqref{eq:gk}. The black dashed curves in (c),(d) and the black dashed circles in (b)-(d) are eye-guiding lines for the well-defined lower-bounded envelope of $S_2^{z}(\mathbf{k}, \omega)$ continuum and the sharpest peak around $\omega \simeq 7$. The momentum cut connects high-symmetry points $\Gamma,\rm M, K, \Gamma$, see also Ref. \cite{re:supp}. All data are normalized by their maximum value.
    }
    \label{fig:csl}
\end{figure}

The behavior of $S_1(\mathbf{k}, \omega) = \sum_{\alpha} S_1^{\alpha} (\mathbf{k},\omega)$ in the CSL phase obtained by the iPEPS ansatz is shown in Fig.~\ref{fig:csl}(a), where the fuzzy continuum signature reflects the fractionalized nature of the CSL, and the gap $\sim 0.2$ (measured in the unit of Kitaev exchange $K_\alpha \equiv 1$) corresponds to the energy gap induced by the two-flux excitation. Note that upon adding $h$, the model is no longer exactly solvable. However, data obtained by iPEPS still are consistent with the leading order perturbative calculation of Ref. \cite{Knolle2014}. 

Having validated the iPEPS algorithm using $S_1$, now we focus on the total dimer correlations $S_2(\mathbf{k}, \omega) = \sum_\alpha S_2^\alpha(\mathbf{k}, \omega)$ and its $z$ component $S_2^{z}(\mathbf{k}, \omega)$ shown in Figs.~\ref{fig:csl}(b,c), which are relevant for high-order processes such as two-spin-flip \cite{Kumar2018,Schlappa2018}.
The most notable feature of $S_2(\mathbf{k}, \omega)$ in Fig.~\ref{fig:csl}(b) is the sharp flat intensity at $\omega \sim 0.5$ which spans the whole Brillouin zone. This resembles that of the flux dynamics in the Abelian phase of the Kitaev model, where the lowest-lying peak is attributed to gapped flux excitations \cite{Knolle2014}.  Indeed, as we will elaborate later, the low-energy peak Fig.~\ref{fig:csl}(b) is attributed to the four-flux excitation. 
In contrast to the total $S_2(\mathbf{k}, \omega)$, the $z$ component of the dimer dynamics $S_2^{z}(\mathbf{k}, \omega)$ exhibits a fractional continuum covering the entire Brillouin zone, as shown in Fig.~\ref{fig:csl}(c). This continuum is marked by prominent intensity peaks at the $\rm K$ point, particularly around $\omega \simeq 7$, as marked in the black dashed circle in Fig.~\ref{fig:csl}(c), which is also visible in the total $S_2(\mathbf{k}, \omega)$ shown in Fig.~\ref{fig:csl}(b);  and the discernible dome-like region at lower energies, as marked by the black dashed curve in Fig.~\ref{fig:csl}(c) where the intensity of $S_2^{z}(\mathbf{k}, \omega)$ becomes weak within the two dome-like envelops. 
The marked contrast between $S_2^{z}(\mathbf{k}, \omega)$ and $S_2(\mathbf{k}, \omega)$ in the CSL arises because, while the contribution from $S_2^{x(y)}(\mathbf{k}, \omega)$ to $S_2(\mathbf{k}, \omega)$ contains flux excitations, the $z$ component $S_2^{z}(\mathbf{k}, \omega)$, as will be discussed in detail later, is virtually only sensitive to the fractionalized Majorana fermions under weak magnetic field, \emph{separating out the fractionalized matter sector from the gauge sector}. Therefore, we note that these discernible features, including the dichotomy between $S_2^{z}(\mathbf{k}, \omega)$ and $S_2(\mathbf{k}, \omega)$, serve as definitive signatures of the CSL of the Kitaev model accessible in scattering experiments.

Given the iPEPS data, we now give an analytical account for the dimer spectra using perturbation theory.  Despite the model losing its integrability at a finite $h$, significant insights can still be extracted from iPEPS through perturbative approximations. 
To the leading order, the dynamical behaviors of $\mathcal{D}_j^{x(y)}$ and $\mathcal{D}_j^z$ in the CSL are distinct.
Note an eigenstate under weak perturbation can be separated into its gauge sector where excitations are $Z_2$ fluxes, and the Majorana sector that is equivalent to a $p+ip$ superconductor \cite{kitaev2006anyons,Read2000,leezhangxiang07,Burnell2011}. 
Hence, $\mathcal{D}_j^{x(y)}$ induces  four fluxes accompanied by the creation of a local Majorana pair $c_j c_{j+z}$ \cite{re:supp}:
\begin{align}
\mathcal{D}_j^{x(y)} \ket{M_{0};\honey{\w}{\w}{\w}{\w}} &= i c_j c_{j+z} \ket{M_{0};\honey{\g}{\g}{\g}{\g}} \label{eq:dxflux}
\end{align}
The ket combines information about the Majorana and the gauge sector; the free Majorana sector $M_0$ is conditioned on a zero-flux configuration in the gauge sector, the white (gray) plaquette denotes the absence (presence) of a $\pi$ flux, and  
the $z$ bond is denoted by the horizontal link.

Given that these fluxes are static excitations in zero or a perturbative magnetic field aligned in the [111] direction, the resultant spectrum is characterized by a flat band of static flux composites. The strongest signal appears at an energy approximately equal to $0.5$, corresponding to the gap induced by a four-fluxes excitation. Consequently, as illustrated in Fig.~\ref{fig:csl}(b), the composite dimer dynamics $S_2(\mathbf{k},\omega)=\sum_{\alpha}S^{\alpha}_2(\mathbf{k},\omega)$ combines the flat flux bands attributable with the fractional continuum emanating from amplitude modes in the free Majorana sector \cite{re:supp}. 

In contrast, $\mathcal{D}_j^z$ does not excite flux, thus the dynamics are solely governed by the Majorana sector:
\begin{align}
\mathcal{D}_j^z \ket{M_{0};\honey{\w}{\w}{\w}{\w}} &= i c_j c_{j+z} \ket{M_{0};\honey{\w}{\w}{\w}{\w}} \label{eq:dxflux1}
\end{align}
The $z$-component of the dimer-dimer correlation $S_2^{z}(\mathbf{k}, \omega)$, being exclusively associated with the Majorana sector and independent of the gauge component, allows for an analytical computation within the framework of a non-interacting $p+ip$ superconducting model within leading order perturbation theory. 
Focusing on the zero-flux sector, 
the spectrum of $S_2^{z}$ can be readily calculated in the Lehmann representation \cite{re:supp}:
\begin{equation} \label{eq:s2zz}
\begin{split}
        S_2^{z} (\mathbf{k},\omega) = \frac{\sqrt{3}}{16 \pi^2} \int_{\rm BZ} G(\mathbf{k}-\mathbf{q}) \delta(\omega - \varepsilon_{\mathbf{k},\mathbf{q}}) d^2\mathbf{q}
\end{split} 
\end{equation}
where the energy of a complex-fermion pair is given by $\varepsilon_{\mathbf{k},\mathbf{q}} \equiv E_{\mathbf{k} - \mathbf{q}} + E_{\mathbf{q}}$, and the spectral weight $G(\mathbf{k}-\mathbf{q})$ is calculated analytically:
\begin{equation} \label{eq:gk}
    G(\mathbf{k} - \mathbf{q}) = \frac{1}{4}\frac{E_{\mathbf{k} - \mathbf{q}}^2}{E_{\mathbf{k} - \mathbf{q}}^2 - Q_{\mathbf{k} - \mathbf{q}}^2}
\end{equation}
We used $E_\mathbf{k}$ to denote the positive fermion band in the flux-free Hamiltonian, and $Q_\mathbf{k} \equiv 4g[\sin(\mathbf{k}\cdot \mathbf{n}_2) - \sin(\mathbf{k}\cdot \mathbf{n}_1) - \sin(\mathbf{k}\cdot (\mathbf{n}_2-\mathbf{n}_1))]$ is due to the next nearest neighbor hopping amplitude $g$, i.e. the leading order time-reversal (TR)-breaking perturbation in the zero-flux sector \cite{kitaev2006anyons}. Our analytical results are shown in Fig.~\ref{fig:csl}(d), which agrees qualitatively with the iPEPS result in  Fig.~\ref{fig:csl}(c).  
Specifically, the sharp spot at $\omega \simeq 7$ and the low-energy envelope in $S_2^{z}(\mathbf{k}, \omega)$ directly reflect the band structure of itinerant Majorana fermions, hence can be particularly useful in the RIXS for singling out the fractionalized Majorana degree of freedom in relevant candidate materials. 
Also note that the weak iPEPS signals below the dashed black curves in Fig.~\ref{fig:csl}(c), which is absent in the analytical result in Fig.~\ref{fig:csl}(d). This is due to induced fluxes for weak fields that are ignored within perturbation theory, where only Majoranas are responsible for the dynamics.

\begin{figure}
    \centering
    \includegraphics[width=\linewidth]{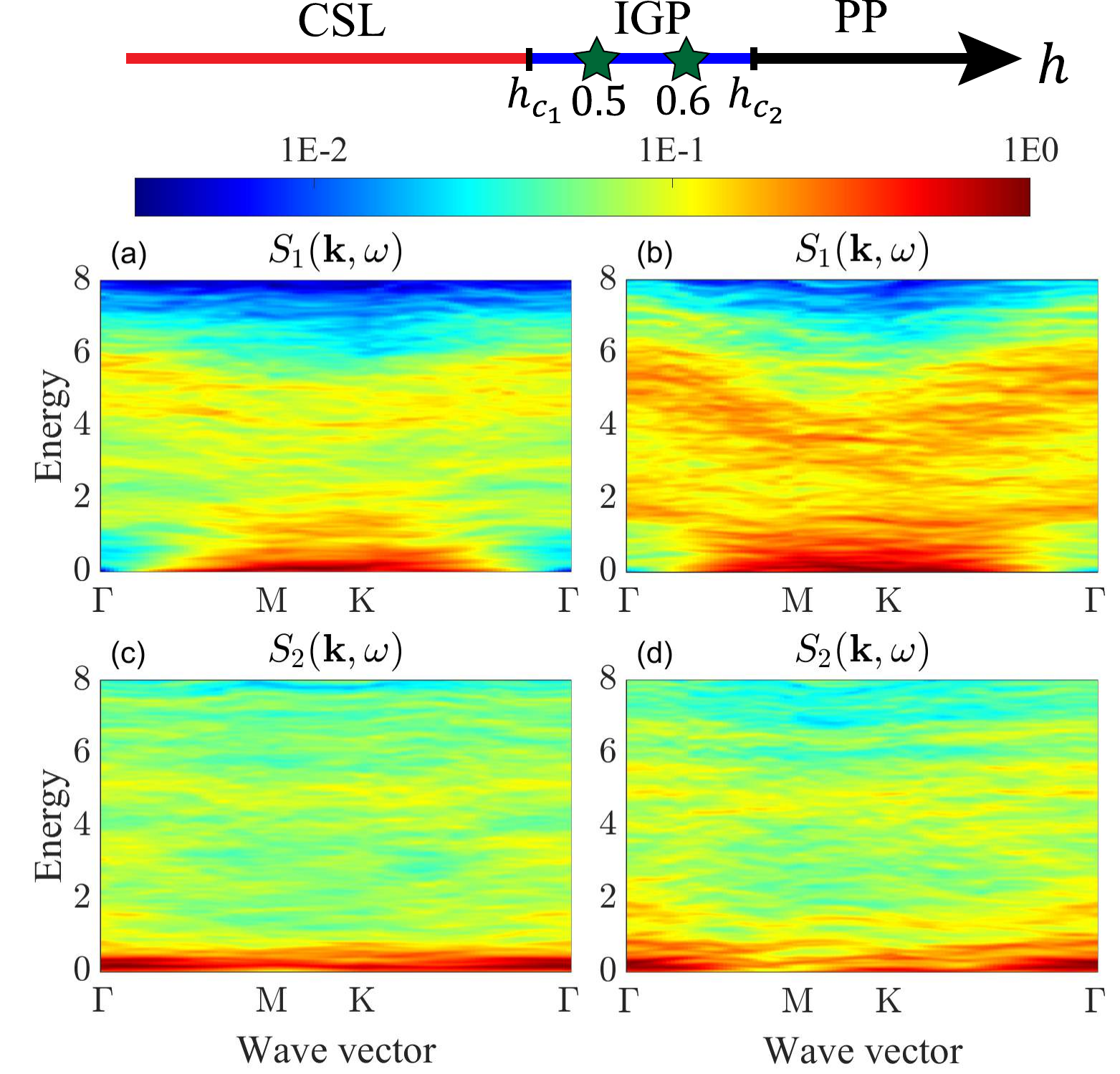}
    \caption{Single- and two-spin flip spectra in IGP presented on a logarithmic color scale along the momentum path $\Gamma \rm M K\Gamma$. (a) Total spin-spin spectrum $S_1(\mathbf{k},\omega)$ at $h=0.5$. (b) $S_1(\mathbf{k},\omega)$ at $h=0.6$. (c) Total dimer-dimer spectrum $S_2(\mathbf{k},\omega)$ at $h=0.5$. (d) $S_2(\mathbf{k},\omega)$ at $h=0.6$. Data are obtained by iPEPS with bond dimension $D = 5$ and a broadening factor $\eta=0.05$.  }
    \label{fig:igp}
\end{figure}

\magenta{\it Correlations in IGP .| }
The most notable findings in IGP at the intermediate field are illustrated in Fig.~\ref{fig:igp}. Two essential signatures of IGP in the single-spin-flip dynamics are observed in  Figs.~\ref{fig:igp}(a),(b).  Noticeably, the spectrum immediately above zero energy is very broad, reflecting its highly fractionalized nature in contrast to those of CSL and PP \cite{re:supp}. 
At low energy, the spectrum is gapless at the $\rm M$ point at a lower field ($h=0.5$), and with an increasing field at both $\rm M$ and $\rm K$ points with $\rm K$ having a stronger signal ($h=0.6$). Such shifting of gapless modes in IGP by $h$ is qualitatively consistent with the previous investigation by classical-shadow tomography where a tunable Friedel-type oscillation was found in the same phase \cite{zhang2023machine}, reflecting its gapless fermionic nature. 

The dimer dynamics shown in Figs.~\ref{fig:igp}(c),(d) further reveal distinguishable gapless signals at low energy, located primarily at $\Gamma$, with noticeable but weaker signal at $\rm K$, and weak or negligible signal near $\rm M$. This is in sharp contrast to the single-flip dynamics of IGP, as well as to the dimer dynamics in the CSL phase. These observations provide robust evidence for the gapless nature of the IGP. Notably, the data derived from iPEPS are free from the constraints of finite size, a limitation often encountered in previous studies utilizing ED and DMRG methods \cite{hickey2019emergence,David2019,Patel12199}. This advantage effectively eliminates the concern of spurious gaplessness that might arise in finite-geometry clusters.

\begin{figure}
    \centering
    \includegraphics[width=\linewidth]{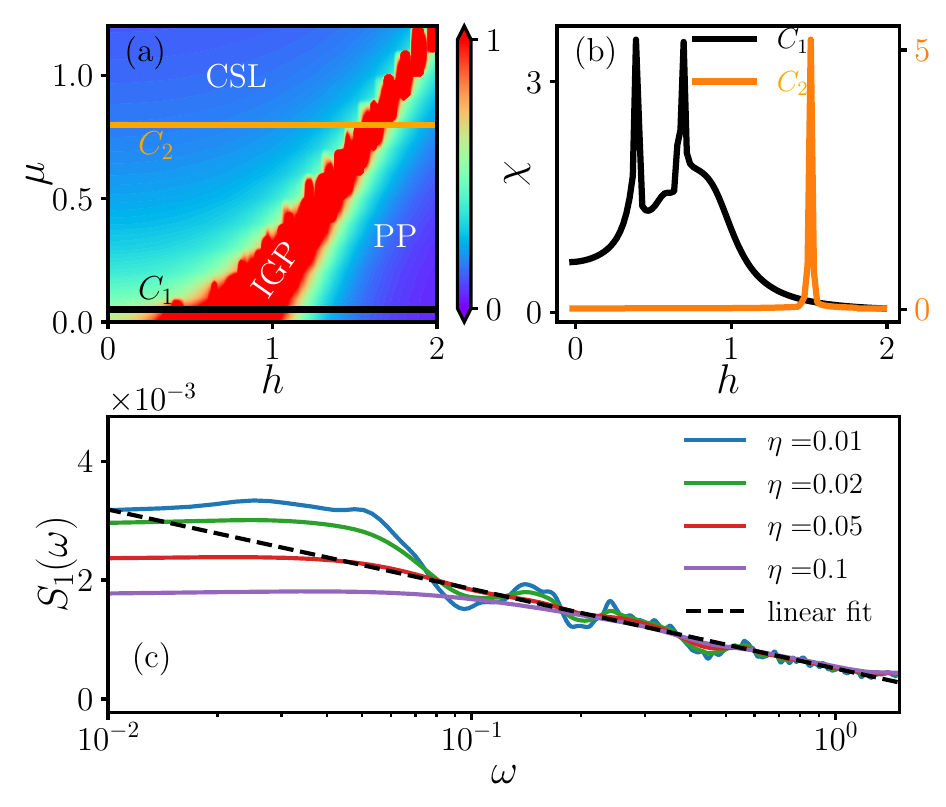}
    \caption{(a) The role of proliferating fluxes in the emergence of the Majorana metal, as evidenced by susceptibility measurements $\chi(\mu,h) = N^{-1} \partial^2 E_{0} /\partial h^2$. The IGP gradually disappears as flux excitation becomes more energetically penalized. Data was obtained by exactly diagonalizing $H'$ in a 24-site cluster ($3\times 4$ unit cells) under PBC.  (b) $C_1$ and $C_2$ cuts in (a). (c) The normalized spectra $S_1(\omega)$ integrated over the first BZ, obtained by iPEPS with varying broadening factor $\eta$ across the second Brillouin zone at $h=0.6$.   }
    \label{fig:chern}
\end{figure}

\magenta{\it  IGP induced by flux fluctuations.|} 
We now discuss the mechanism for the emergence of IGP. Within IGP, the ground state is no longer flux-free. Instead, quantum fluctuations of the flux become significant, leading to the nucleation of $\pi$-fluxes. We argue that the fluxes fluctuations are vital to the mechanism that drives the gaplessness of the intermediate phase. 
As a test, we first validate its contrapositional statement: that if flux fluctuations are energetically suppressed, IGP would be consequently removed. We therefore modify the original Hamiltonian to include an additional energy penalty for flux excitations: $H' = \sum_{\langle ij\rangle, \alpha}{K_\alpha \sigma^\alpha_i \sigma^\alpha_j} - \mu \sum_p W_p - h\sum_{i,\alpha}\sigma_i^\alpha$. 
In this formulation, the term $- \mu \sum_p W_p$ with $\mu > 0$ renders flux excitations energetically unfavorable. The summation $\sum_p W_p$ commutes with the pure Kitaev honeycomb model, consequently, it penalizes flux excitations without influencing the dynamics of the itinerant Majorana fermions. This allows for a direct examination of the role of flux fluctuations in the emergence and characteristics of IGP. 
Figures~\ref{fig:chern}(a),(b) show the magnetic susceptibility of $H'$ obtained by 24-site ED under PBC. 
For $\mu \simeq 0$ obtained by ED, the intermediate phase, corresponding to IGP \cite{hickey2019emergence} emerges under a finite magnetic field and persists for a finite range of $h$ before the confinement transition into PP. However, as demonstrated in Figs.~\ref{fig:chern}(a),(b), at larger $\mu$ whereby the flux fluctuation is suppressed, the IGP spans a smaller range of $h$ until vanishing at $\mu \gtrsim 0.6$. This showcases the important role played by the finite flux density in the formation of IGP; and suggests that a faithful effective theory thereof must consist of two dominant fields: itinerant Majorana fermions and localized fluxes, which cannot be captured by solving quadratic parton self-consistent equations \cite{Jiang2020,ZhangNatComm2022}, since a flux excitation is a many-body entangled state of the $Z_2$ field, or equivalently, of the bond fermions \cite{Hong10,feng2023stat}. 

To further establish the connection between the IGP and Majoranas, we note that the Majorana fermion sector of the Kitaev model is depicted as a Majorana-hopping model of class~D, which are known to have three phases: a topological insulator, a trivial insulator, and a gapless metal phase~\cite{Senthil2000,Chalker2001,Huse2012,Knolle2019}. In the Kitaev model under a magnetic field, the topological insulator phase corresponds to CSL, and the gapless metal phase can arise from the fluctuating flux configurations induced by magnetic field~\cite{zhu2024emergent}, thus
the IGP is effectively a Majorana metal phase, with a characteristic $\log E$ scaling of the density of states (DOS) near zero \cite{Senthil2000,Huse2012}.  
At the microscopic level, such scaling of DOS can be tested by the spectrum of $S_1(\omega)$, whose low-energy dynamics are primarily attributed to metallic Majorana fermions. This is because 
fluxes are virtually unseen by the two-point correlation, i.e. that the two local flips of flux do not significantly affect the averaged Majorana band conditioned on an exponentially large ensemble of proliferating flux configurations.  Importantly, the logarithmic scaling of DOS is robustly validated by iPEPS results for various broadening factors $\eta$, as shown in Fig.~\ref{fig:chern}(c), providing direct support for identifying the IGP phase as a Majorana metal. 
Notably, with the smallest $\eta$ we still find no observable gap nor the trend of opening a gap in $S_1(\omega)$ at the energy scale $0.01$, which is comparable to or lower than previous putative gaps obtained by parton mean field theories \cite{Jiang2020,ZhangNatComm2022}. The mean field theory does not account for gauge fluctuations. Consequently, the mean-field predictions presented in Refs.~\cite{Jiang2020,ZhangNatComm2022} differ from ours. 
Furthermore, the oscillation pattern under the finest energy resolution, shown in Fig.~\ref{fig:chern}(c), and the plateau at the lowest energies are in good agreement with the metal phase of the class~D predicted by random matrix theory \cite{Altland97,Huse2012}.

\magenta{\it Conclusion and Outlook .|}
In this work, we have elucidated the signatures of fractionalization in Kitaev QSLs as a function of a  [111] magnetic field with a focus on the chiral spin liquid (CSL) and the intermediate gapless Majorana metal phase (IGP). Utilizing iPEPS and analytical methods, we have identified dynamical signatures of fractionalized quasi-particles through spin-spin and dimer-dimer correlations. As the community awaits definitive fingerprints of QSLs, there is a need to develop new classes of experiments closely guided by theory. We believe our predictions here can provide the necessary impetus for measuring higher-order dynamical spin correlations using INS, RIXS, and pump-probe spectroscopy.


\begin{acknowledgments}
\textbf{Acknowledgments: }
K.W. and T.X. are grateful to Bruce Normand, Yi Zhou, Xing-Yu Zhang, Chenguang Liang, Tong Liu, Jia-Lin Chen, Jiahang Hu and Wen-Tao Xu for useful discussions. S.F. and N.T. thank Adhip Agarwala and Subhro Bhattacharjee for their insightful discussion and comments. 
This work is supported by the National Key Research and Development Project of China (Grants No.~2021ZD0301800 and No.~2022YFA1403900), the National Natural Science Foundation of China (Grants No.~12488201, No.~11874095, No.~11974396, and Grants No.~12347107, No.~12322403), the Strategic Priority Research Program of Chinese Academy of Sciences (Grants No.~XDB0500202, ~XDB33010100 and No.~XDB33020300) and the Youth Innovation Promotion Association of Chinese Academy of Sciences (Grant No.~2021004).
S.F., P.Z. are funded by the U.S. National Science Foundation's Materials Research Science and Engineering Center under Award No. DMR-2011876 and N.T. by Award No. DMR-2138905. 
\end{acknowledgments}

\bibliography{spectra}

\appendix
\begin{widetext}
\clearpage



\begin{center}
\textbf{\large Supplemental Material for \\ ``
Fractionalization Signatures in the Dynamics of Quantum Spin Liquids"}
\end{center}





\setcounter{section}{0}
\setcounter{figure}{0}
\setcounter{equation}{0}
\renewcommand{\thefigure}{S\arabic{figure}}
\renewcommand{\theequation}{S\arabic{equation}}
\renewcommand{\thesection}{S\arabic{section}}
\twocolumngrid

\section{iPEPS algorithm}
Tensor Network serves as a highly effective numerical tool for investigating strongly correlated systems, evolving from the well-known density matrix renormalization group (DMRG) algorithm \cite{white1992density}. Over the past decades, various ground state ansatzes have been proposed, including Matrix Product State (MPS) \cite{ostlund1995thermodynamic}, Projected Entangled Pair States (PEPS) \cite{verstraete2004renormalization}, and projected entangled simplex states (PESS) \cite{xie2014tensor}. The first and latter two ansatz adhere to the area law of entanglement in one and two dimensions \cite{cirac2021matrix}, respectively. Beyond the description of ground state properties, numerous algorithms based on tensor networks have been developed to explore dynamical information. The fundamental approach for calculating the dynamical spectrum involves time evolution, employing methods such as Time-Evolving Block Decimation (TEBD) \cite{vidal2004efficient,daley2004time,white2004real} or Time-Dependent Variational Principle (TDVP) \cite{haegeman2011time,haegeman2013post,gohlke2018dynamical,Yang2020,tian2021matrix,wang2024spectral,CBETDVP2024}. However, real-time evolution often faces challenges associated with the volume law, limiting the duration of evolution and resulting in the loss of low-frequency information. Other algorithms, such as Lanczos and Chebyshev methods \cite{hallberg1995density,kuhner1999dynamical,holzner2011chebyshev,xie2018reorthonormalization,xiang2023density}, directly address this issue in the frequency domain. These approaches rely on multiple DMRG calculations and typically perform well for finite systems. However, the strong finite-size effects thereof can give rise to spurious modes which vanish in the infinite limit.

In addition to the aforementioned approach, an alternative method involves adopting the single-mode approximation \cite{feynman1954atomic}, which was originally proposed to elucidate the low-lying excitations of superfluids. Ostlund and Rommer~\cite{ostlund1995thermodynamic,haegeman2012variational} introduced this ansatz within the matrix product state framework, and Vanderstraeten {\it et al.}~\cite{vanderstraeten2015excitations} extended it to the PEPS context in tensor network methods. This method can be applied to infinite system and has been proven to be highly effective in computing the dynamical spectrum both in magnetic order~\cite{vanderstraeten2019simulating,ponsioen2020excitations,ponsioen2022automatic,chi2022spin} and spin liquid systems~\cite{tan2023gauge}.

The iPEPS-based single-mode approach enables the computation of high-resolution, momentum-resolved dynamics in the thermodynamic limit, significantly outperforming traditional numerical methods. Exact Diagonalization (ED), limited by very small system sizes, faces difficulties in detecting gapless excitations due to finite size effects, compounded by its poor momentum resolution. The aforementioned dmrg-based algorithms can handle dynamics for larger systems, including multiple chains, yet are still restricted by the size of the system. Quantum Monte Carlo (QMC) is widely applied in the study of strongly correlated systems; however, it faces significant challenges of the sign problem, particularly in frustrated or fermionic systems. Additionally, its reliance on imaginary time evolution introduces the issue of analytic continuation in computing dynamics, which is a notoriously difficult problem. Linear spin wave theory (LSWT) accounts for quantum corrections to classical magnetic order and serves as a universal method for calculating low-energy excitations in quantum magnets. However, it falls short when addressing quantum spin liquid (QSL) systems and is unable to compute high-energy excitations.

The Kitaev model discussed in the main text is defined on the honeycomb lattice with different interactions on three bonds, as illustrated in Fig.~\ref{Ansatz} (a). There are two sites and three bonds in one unit cell. To facilitate network contraction, we initially merge the $A$ and $B$ sites, transforming the lattice into a square lattice. Subsequently, we define the local tensor of iPEPS on the square lattice, as depicted in Fig.~\ref{Ansatz} (c). This allows us to implement the corner transfer matrix renormalization group (CTMRG)\cite{CTMRG96,Orus_CTM09,Corboz_tJ14} method for network contraction and measurement of observables. We store the boundary tensors and projectors of the ground states to avoid the necessity for singular value decomposition (SVD) in the computation of excitations, thereby facilitating GPU acceleration. Furthermore, we adopt a fixed-point methodology in the implementation of CTMRG\cite{liao2019differentiable,ponsioen2022automatic} to minimize the number of backward steps and significantly reduce computational expenses. The maximum bond dimension of the environment is controlled by $\lambda$. The iPEPS ansatz is initially optimized using the simple update method \cite{jiang2008accurate}, followed by further optimization through energy minimization using automatic differentiation techniques \cite{liao2019differentiable}.


To characterize the excited states, we employ the single-mode approximation~\cite{feynman1954atomic}, which was originally proposed to elucidate the low-lying excitations of superfluids, yields a variational ansatz for the excited state
\begin{equation}
\ket{\Psi_\mathbf{k}}=\sum_{\mathbf{r}}e^{-i\mathbf{k}\cdot\mathbf{r}}\ket{\Psi_\mathbf{r}}
\end{equation}
where $\mathbf{k}$ denotes momentum and $\ket{\Psi_\mathbf{r}}$ is the state with an excitation at site $\mathbf{r}$. 
In these methods, $\mathbf{r}$ is represented by substituting the local tensor $A$ of the ground state at site $\mathbf{r}$ with a perturbed local tensor $B$.  The single-mode ansatz is illustrated in Fig.~\ref{Ansatz}(d).

\begin{figure}[t] 
\centering 
\includegraphics[width=0.45\textwidth]{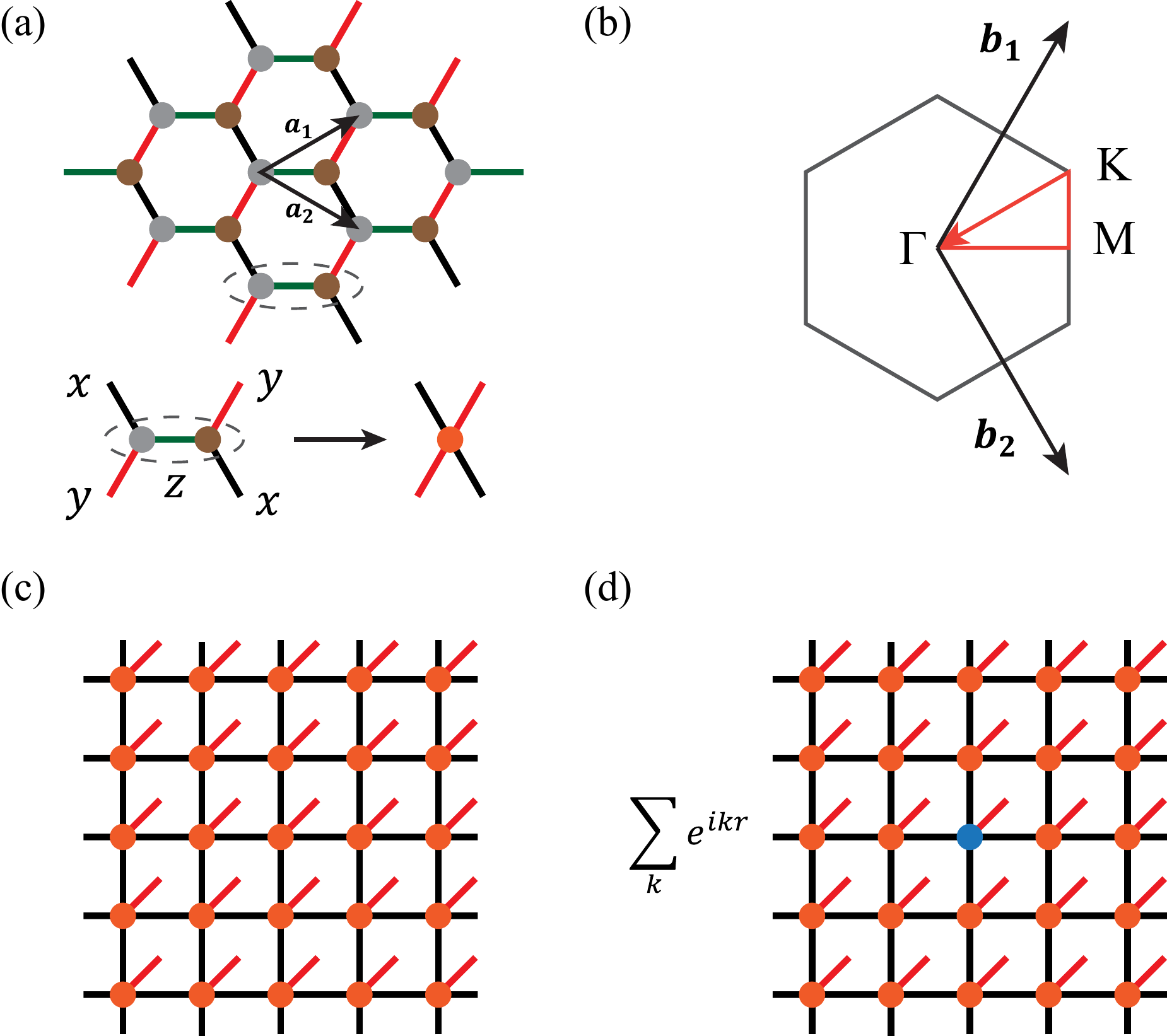} 
\caption{(a) Kitaev model on honeycomb lattice and transformation to the square lattice by merging two sites. (b)  The first Brillouin zone and the momentum path $\Gamma-\rm M-K-\Gamma$. (c) Ground state iPEPS Ansatz defined on the square lattice, the local physical bond dimension is 4.  (d) Excited state Ansatz.} 
\label{Ansatz} 
\end{figure}

The excited states must fulfill the orthogonality constraint concerning the ground state $\ket{\psi_0}$:
\begin{equation}
\braket{\psi_0}{\Psi_\mathbf{k}(B)}=0
\end{equation}
For $k\neq0$, the constraint is automatically satisfied due to momentum conservation. However, when $k=0$, the variational space should exclude the ground state:

\begin{equation}
\braket{\psi_0}{\Psi_{k=0}}=N\braket{\psi_0}{\Psi_r}=N (M\cdot B) = 0
\end{equation}
where $M$ denotes the contracted tensor of the entire tensor network excluding the $B$ tensor, obtained through the CTMRG method. 
In this context, the overlap can be expressed as the contraction of two layers of the tensor network. Importantly, $B$ must be confined within the subspace orthogonal to $M$. 

In addition to the aforementioned constraint, it is imperative to establish a fixed gauge for the PEPS. The excited state ansatz exhibits invariance under the gauge transformation:
$$B \to  B+e^{-i\mathbf{k}}AG-GA$$
This transformation is visually depicted as:
\begin{figure}[h] 
\centering 
\includegraphics[width=0.45\textwidth]{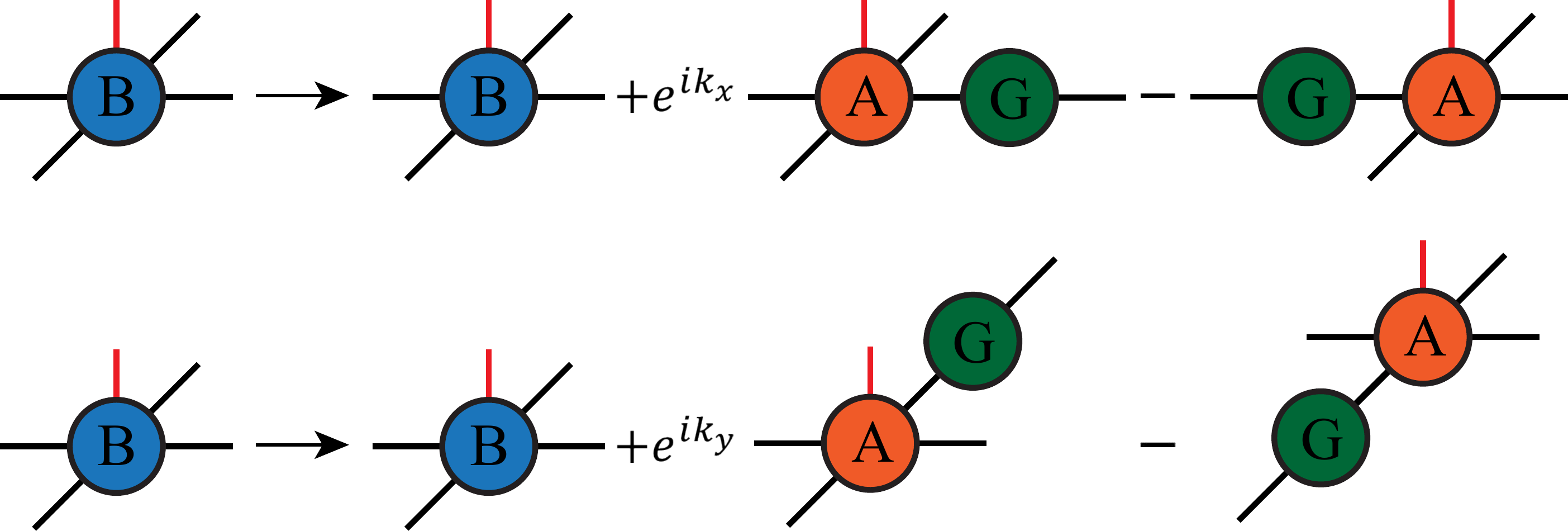} 
\caption{Gauge transformation along the $x$ and $y$ axis.} 
\end{figure}

We define $B_X = e^{ik_x}AG-GA$ and $B_Y = e^{ik_y}AG-GA$. If we were to set $B = B_X$ or $B_Y$, this would yield a null state. Therefore, to ensure physically viable states, we must confine the variational space to exclude the $B_X$ and $B_Y$ tensor spaces. Combining these restrictions, the physically allowed $B$ tensors must satisfy 
\begin{equation}
(M,B_X,B_Y)^\dagger \cdot B = 0
\end{equation}
The constraints are visually depicted in Fig.~\ref{Fig:S3}.
It is worth noting that the rank of $G$ is $D^2$. Consequently, the overall rank of the matrix is at most $2D^2 + 1$. As a result, there exist at most $2D^2 + 1$ linearly independent vectorized $B$ tensors that need to be excluded in the tangent space. Given that the dimension of $B$ is $dD^4$, we can infer that there are at least $N_B=dD^4 - 2D^2 - 1$ basic solutions, denoted as $\hat{B}_n$ ($n=1, \dots, N_B$).

\begin{figure}[h] 
\centering
\includegraphics[width=0.45\textwidth]{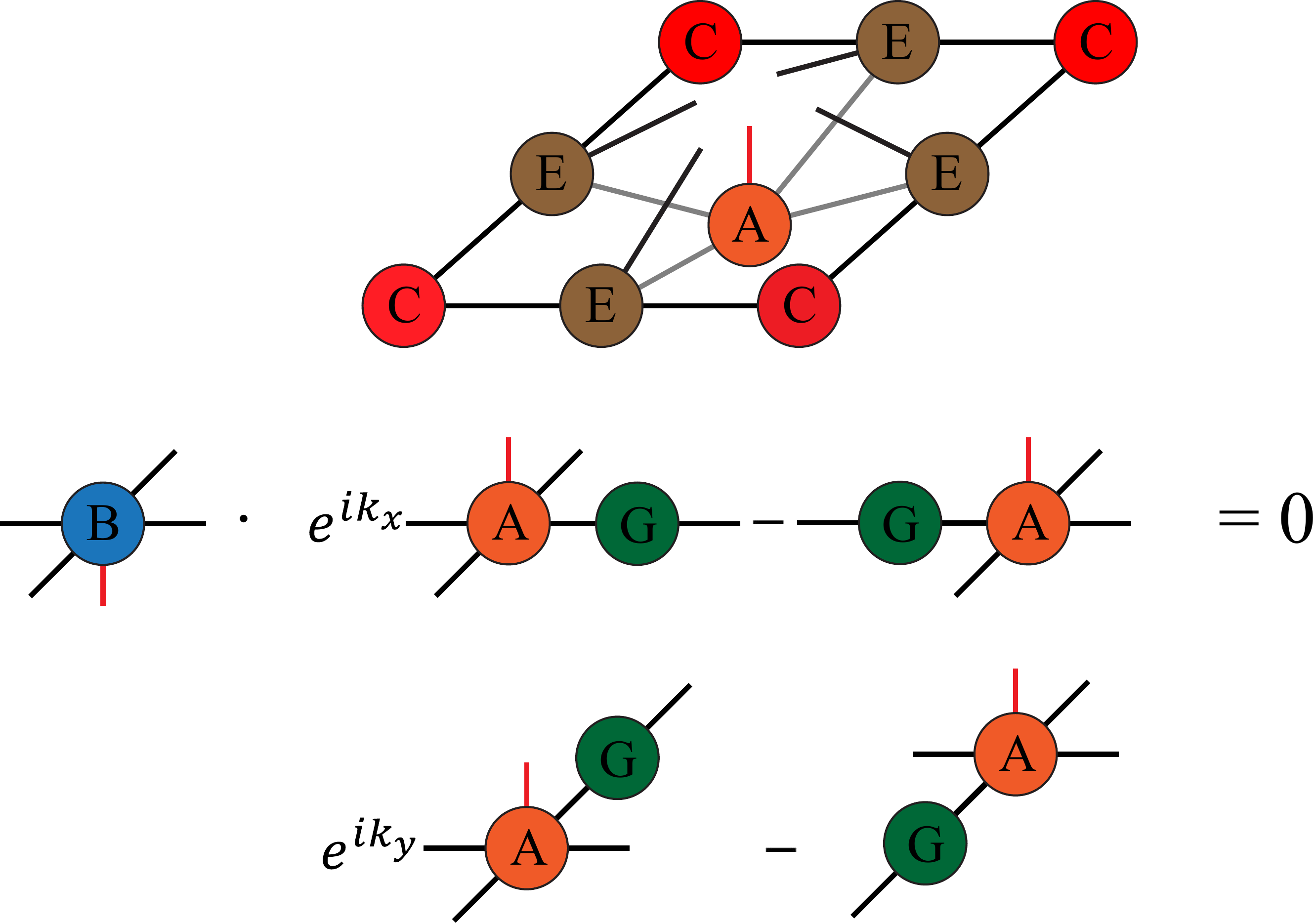} 
\caption{Graphical representation of the orthogonality conditions. $C$ and $E$ are corner and edge tensors in CTMRG algorithm, respectively.}
\label{Fig:S3}
\end{figure}

To target excited states, we need to minimize the cost function
\begin{equation}
    L = {\frac{\bra{\Psi_\mathbf{k}(B)}H-E_{\rm gs}\ket{\Psi_\mathbf{k}(B)}}{\braket{\Psi_\mathbf{k}(B)}{\Psi_\mathbf{k}(B)}}}
\end{equation}
to obatin $B$ tensor. The minimization equals to solving the equation 
\begin{equation}
    \frac{\partial}{\partial B}{\bra{\Psi_\mathbf{k}(B)}H-E_{\rm gs}\ket{\Psi_\mathbf{k}(B)}}=E_\mathbf{k}\frac{\partial}{\partial B}{\braket{\Psi_\mathbf{k}(B)}{\Psi_\mathbf{k}(B)}}
\label{min}
\end{equation}
where $E_\mathbf{k}$ is the excitation energy. The dimension of $B$ tensor is $dD^4$. We can parameterize $B$ with the vectorized basis obtained above
\begin{align}
    B = \sum_n b_n \hat{B}_n \quad (n=1,\dots,N_B)
\end{align}
Then we define the effective Hamiltonian $H_\mathbf{k}$ and the norm matrix $N_\mathbf{k}$ with elements written as
\begin{equation}
\begin{aligned}
    {\bra{\Psi_\mathbf{k}(\hat{B}_m)}H-E_{\rm gs}\ket{\Psi_\mathbf{k}(\hat{B}_n)}}=\hat{B}_m^\dagger H_\mathbf{k}^{mn} \hat{B}_n\\
    {\braket{\Psi_\mathbf{k}(\hat{B}_m)}{\Psi_\mathbf{k}(\hat{B}_n)}}=\hat{B}_m^\dagger N_\mathbf{k}^{mn} \hat{B}_n
\end{aligned}
\end{equation}
Substituting into Eq.~\eqref{min}, the equation becomes 
\begin{equation}
    H_\mathbf{k} b = E_\mathbf{k} N_\mathbf{k} b
\end{equation}
where $b=(b_1,\dots,b_{N_B})^T$. By solving this general eigen equation, the local tensor $B$ can be determined.

The zero-temperature dynamical spectral function is defined as
\begin{equation}
\begin{aligned}
\label{spec}
    &S^{\alpha\beta}(\mathbf{k},\omega)=\bra{\psi_0}O^\alpha_{\mathbf{k}} \delta(\omega-H+E_{\rm gs}) O^\beta_{-\mathbf{k}}\ket{\psi_0}\\
    &=\sum_m{\bra{\psi_0}O^\alpha_\mathbf{k}\ket{\Psi_\mathbf{k}^m}\bra{\Psi_\mathbf{k}^m}O^\beta_{-\mathbf{k}}\ket{\psi_0}}\delta(\omega-E_\mathbf{k}^m+E_{\rm gs}) 
\end{aligned}
\end{equation}
where $\alpha,\beta=x,y,z$. $\ket{\psi_m}$ denotes the eigenstate of hamiltonian $H$ with energy $E_m$. The corresponding spectral weight can be obtained by contracting the double-layer tensor
\begin{equation}
\begin{split}
\bra{\Psi_\mathbf{k}^m}O^\alpha_\mathbf{k}\ket{\psi_0}&=\frac{1}{\sqrt{N}}\sum_{\mathbf{rr'}}e^{-i\mathbf{k\cdot(r-r')}}\bra{\Psi_\mathbf{r'}^m}O^\alpha_\mathbf{r}\ket{\psi_0}.
\label{weight}
\end{split}
\end{equation}
The contraction can be obtained using the CTM summation introduced in Ref.~\cite{ponsioen2022automatic}. The delta function in Eq.~\eqref{spec} can be approximated using the Lorentzian expansion with a broadening factor $\eta$.

In the main text, we have focused on the spectrum of two kinds of excitations: one spin flip $\sigma_i^\alpha$ and two spin flip $\sigma_i^\alpha \sigma_{i+z}^\alpha (\alpha = x,y,z)$. We calculated the spectrum function in momentum space 
\begin{equation}
\begin{split}
S^{\alpha}_1(\mathbf{k},\omega)&=\frac{1}{N}\sum_{ij} e^{-i\mathbf{k
\cdot(R_i-R_j)}}[S_{aa}^{\alpha}(i,j,\omega)+S_{bb}^{\alpha}(i,j,\omega)\\&+e^{-i\mathbf{kr_{ab}}}S_{ab}^{\alpha}(i,j,\omega)+e^{i\mathbf{kr_{ab}}}S_{ba}^{\alpha}(i,j,\omega)]\\
S^{\alpha}_2(\mathbf{k},\omega)&=\frac{1}{N}\sum_{ij} e^{-i\mathbf{k\cdot(R_i-R_j)}}S_{2}^{\alpha}(i,j,\omega)
\end{split}
\end{equation}
where $\mathbf{R_i}$ and $\mathbf{R_i}$ represent the sites of unit cells, and $\mathbf{r_{ab}}$ is the distance between sites a and b within one unit cell. $S_{a/b,a/b}^\alpha(i,j,\omega)$ and $S_{2}^\alpha(i,j,\omega)$ denote the spectral function in real space. In the Lehmann representation, we have:
\begin{equation}
\begin{split}
S_{a/b,a/b}^\alpha(i,j,\omega)=\sum_m &\bra{\psi_0}\hat{\sigma}_{i,a/b}^\alpha\ket{\Psi_\mathbf{k}^m}\bra{\Psi_\mathbf{k}^m}\hat{\sigma}_{j,a/b}^\alpha\ket{\psi_0}\\
&\times \delta(\omega-E_\mathbf{k}^m+E_{\rm gs})\\
\end{split}
\end{equation}
and
\begin{equation}
\begin{split}
S_{2}^\alpha(i,j,\omega)=\sum_m &\bra{\psi_0}\hat{\sigma}_{i,a}^\alpha\hat{\sigma}_{i,b}^\alpha\ket{\Psi_\mathbf{k}^m}\bra{\Psi_\mathbf{k}^m}\hat{\sigma}_{j,a}^\alpha\hat{\sigma}_{j,b}^\alpha\ket{\psi_0}\\
        &\times \delta(\omega-E_\mathbf{k}^m+E_{\rm gs}) 
\end{split}
\end{equation}
where $\hat{\sigma}_{i,a/b}^\alpha=\sigma_{i,a/b}^\alpha-\bra{\psi_0}\sigma_{i,a/b}^\alpha\ket{\psi_0}$ and $\hat{\sigma}_{i,a}^\alpha\hat{ \sigma}_{i,b}^\alpha=\sigma_{i,a}^\alpha\sigma_{i,b}^\alpha-\bra{\psi_0}\sigma_{i,a}^\alpha\sigma_{i,b}^\alpha\ket{\psi_0}$. By representing the excited state as a summation, we can utilize CTM summation to calculate spectral weight as discussed in Eq.~\eqref{weight}.

\section{Spin-spin correlation}
The spin-spin correlation in the pure Kitaev QSL or CSL under perturbation can be made explicit if we separate an eigenstate into gauge and matter sectors \cite{kitaev2006anyons,Baskaran2007}. For a fixed gauge configuration, the ground state wavefunction can be written as $\ket{\psi} = \ket{M_\mathcal{G},\mathcal{G}}$ with $\mathcal{G}$ denoting the $Z_2$ gauge configuration and  $M_\mathcal{G}$ the matter Majorana fermions on the gauge background. In such representation spin are fractionalized into Majoranas $\sigma_j^a = ic_j b_j^a$, and the Hamiltonian in a particular $\ket{\mathcal{G}}$ sector becomes quadratic and integrable as  $H = i \sum_{\expval{ij}_a} J_a\: u_{\expval{ij}_a} c_i c_j$, where $u_{\expval{ij}_a} = \pm 1$ are good quantum numbers that determine $\ket{\mathcal{G}}$ by pinning down a particular configuration of gauge fluxes $W_p = \pm 1$. In order to explicate useful properties of correlation between spins, we define bond fermions $\eta_{\expval{ij}_a} = \frac{1}{2}(b_i^a + ib_j^a)$ for $i \in A$ and  $j \in B$ sublattice respectively, such that
\begin{equation} \label{eq:bondf}
	\sigma_i^a = ic_i(\eta_{\expval{ij}_a} + \eta_{\expval{ij}_a}^\dagger),\;\;\; \sigma_j^a = c_j (\eta_{\expval{ij}_a}- \eta_{\expval{ij}_a}^\dagger)
\end{equation}
Local spin operators involve only two-point pauli matrices that share a link, each of which can be written as $\sigma_j^a \propto c_j \hat{\pi}_{1,\expval{jk}_a}\hat{\pi}_{2,\expval{jk}_a}$ using bond fermions representation defined in Eq.~\eqref{eq:bondf}, where $\hat{\pi}_{1,\expval{jk}_a}$ and $\hat{\pi}_{2,\expval{jk}_a}$ flip a pair of adjacent fluxes, denoted by subscripts $1,2$, that share the same link $\expval{jk}_a$. It is then readily to see that Pauli spin operators change the matter and gauge sectors of an eigenstate $\ket{n}$ according to:
\begin{align} 
	 \sigma_{j}^x \ket{M_{0};\honey{\w}{\w}{\w}{\w}} &\propto c_j \ket{M_{0};\honey{\g}{\w}{\g}{\w}} \\
     \sigma_{j}^y \ket{M_{0};\honey{\w}{\w}{\w}{\w}} &\propto c_j \ket{M_{0};\honey{\g}{\g}{\w}{\w}}\\
    \sigma_{j}^z \ket{M_{0};\honey{\w}{\w}{\w}{\w}} &\propto c_j \ket{M_{0}; \honey{\w}{\g}{\g}{\w}}
\end{align}
where we have separated the Majorana fermion $c_j$ from the $Z_2$ gauge field. Here $M_0$ denotes the free Majorana sector conditioned on the zero-flux gauge sector. In the gauge sector, flipped fluxes $\Delta \pi = -1$ are denoted by hexagons in light gray, in contrast to the rest of hexagons in white that denote the original flux configuration of $\ket{n}$, and Bravais lattice label $j$ denotes the primitive cell on the central horizontal link.  
By the orthogonality between flux configurations, 
it becomes readily to see that the two-point correlation in the ground state sector of the isotropic Kitaev model is determined by $\langle \sigma_j^\alpha \sigma_{j+\beta}^\alpha\rangle\delta_{\alpha,\beta} \simeq -0.52$ \cite{Baskaran2007,Feng2022}. This remains true for all flux bias $\mu$ defined in the main text since the magnitude of the two-spin correlation only involves the Majorana sector, which remains untouched by $\mu W_p$.

\section{Dimer-dimer correlation}
Here we use the Majorana formalism to calculate the correlation between two z-dimers in the pure Kitaev QSL or the CSL phase of the Kitaev honeycomb model. 
In the flux-free sector, we write the pure Majorana Hamiltonian:
\begin{equation} 
	H_0 = \frac{1}{2}\sum_{\mathbf{q}}
	\begin{pmatrix}
		a_{-\mathbf{q}} & b_{-\mathbf{q}}
	\end{pmatrix}
	\begin{pmatrix}
		Q(\mathbf{q}) & if(\mathbf{q}) \\
		-i f^*(\mathbf{q}) & -Q(\mathbf{q})
	\end{pmatrix}
	\begin{pmatrix}
		a_{\mathbf{q}} \\ b_{\mathbf{q}}
	\end{pmatrix}
\end{equation}
where $a$ and $b$ are momentum representation of Majorana operators $c_{i,A(B)}$ on sublattice $A(B)$:
\begin{equation} \label{eq:hamapp}
    a_{\mathbf{q}} = \frac{1}{\sqrt{2N}}\sum_j e^{-i\mathbf{k}\cdot \mathbf{r}_j} c_{j,A},~b_{\mathbf{q}} = \frac{1}{\sqrt{2N}}\sum_j e^{-i\mathbf{k}\cdot \mathbf{r}_j} c_{i,B}.
\end{equation}
Note that the spectrum is given by the eigenvalues of the matrix $
	\begin{pmatrix}
		Q(\mathbf{q}) & if(\mathbf{q}) \\
		-i f^*(\mathbf{q}) & -Q(\mathbf{q})
	\end{pmatrix}$ instead of half of it. The $1/2$ factor is eliminated due to the redundancy $a_{-\mathbf{q}}=a^{\dag}_{\mathbf{q}}$ and $b_{-\mathbf{q}}=b^{\dag}_{\mathbf{q}}$.
The time-reversal (TR) symmetry is broken due to
\begin{equation}
    Q(\mathbf{k}) = 4g[\sin(\mathbf{k}\cdot \mathbf{n}_2) - \sin(\mathbf{k}\cdot \mathbf{n}_1) - \sin(\mathbf{k}\cdot \mathbf{n}_3)] 
\end{equation}
where $\mathbf{n}_3 = \mathbf{n}_2-\mathbf{n}_1$. 
The off-diagonal elements for each mode is related to $f(\mathbf{q}) =2(K_x e^{i\mathbf{q} \cdot \mathbf{n}_1} + K_y e^{i\mathbf{q}\cdot \mathbf{n}_2} + K_z) \equiv  2(K_x e^{iq_x} + K_y e^{iq_y} + K_z) $ where we've defined $q_x \equiv \mathbf{q}\cdot \mathbf{n}_1$ and $q_y \equiv \mathbf{q}\cdot \mathbf{n}_2$. For the convenience of derivation, we split its real and imaginary parts into \cite{Yang07,Feng2022}:
\begin{align}
    f(\mathbf{q}) &= \epsilon_\mathbf{q} + i\Delta_\mathbf{q} \\
    \epsilon_\mathbf{q} &= 2(K_x \cos q_x + K_y\cos q_y + K_z),\\
\Delta_\mathbf{q} &= 2(K_x\sin q_x + K_y \sin q_y)
\end{align}
We investigate the dimer dynamics in the TR-breaking case. 
In the diagonal basis of the complex fermions $C$ we have
\begin{equation}
\begin{split}
    H_{0} = \sum_{\mathbf{k}} E_\mathbf{k} \left( C_{\mathbf{k},1}^\dagger C_{\mathbf{k},1} - C_{\mathbf{k},2}^\dagger C_{\mathbf{k},2} \right) 
\end{split}
\end{equation}
The energy is given by
\begin{equation}
    E_\mathbf{q} = \pm \sqrt{Q_\mathbf{q}^2 + \abs{f_\mathbf{q}}^2} = \pm \sqrt{Q_\mathbf{q}^2 + \Delta_\mathbf{q}^2 + \epsilon_\mathbf{q}^2}
\end{equation} 
with the ground state given by filling the negative band of $C_2$ fermion $\ket{\rm gs} = \prod_\mathbf{k} C_{\mathbf{k},2}^\dagger\ket{0}$, 
where the complex fermion modes created by $C_2^\dagger$ are related to Majoranas by 
\begin{equation}
\begin{split}
    a_\mathbf{q} &= \frac{\Delta_\mathbf{q} - i\epsilon_\mathbf{q}}{\sqrt{2}(E_\mathbf{q}-Q_\mathbf{q})} \left( C_{\mathbf{q},2} - C_{-\mathbf{q},2}^\dagger \right), \\
    b_\mathbf{q} &= \frac{1}{\sqrt{2}}\left( C_{\mathbf{q},2} + C_{-\mathbf{q},2}^\dagger \right) \label{eq:aqbq}
\end{split}
\end{equation}
\begin{figure*}[t]
\centering
\includegraphics[width=0.95\textwidth]{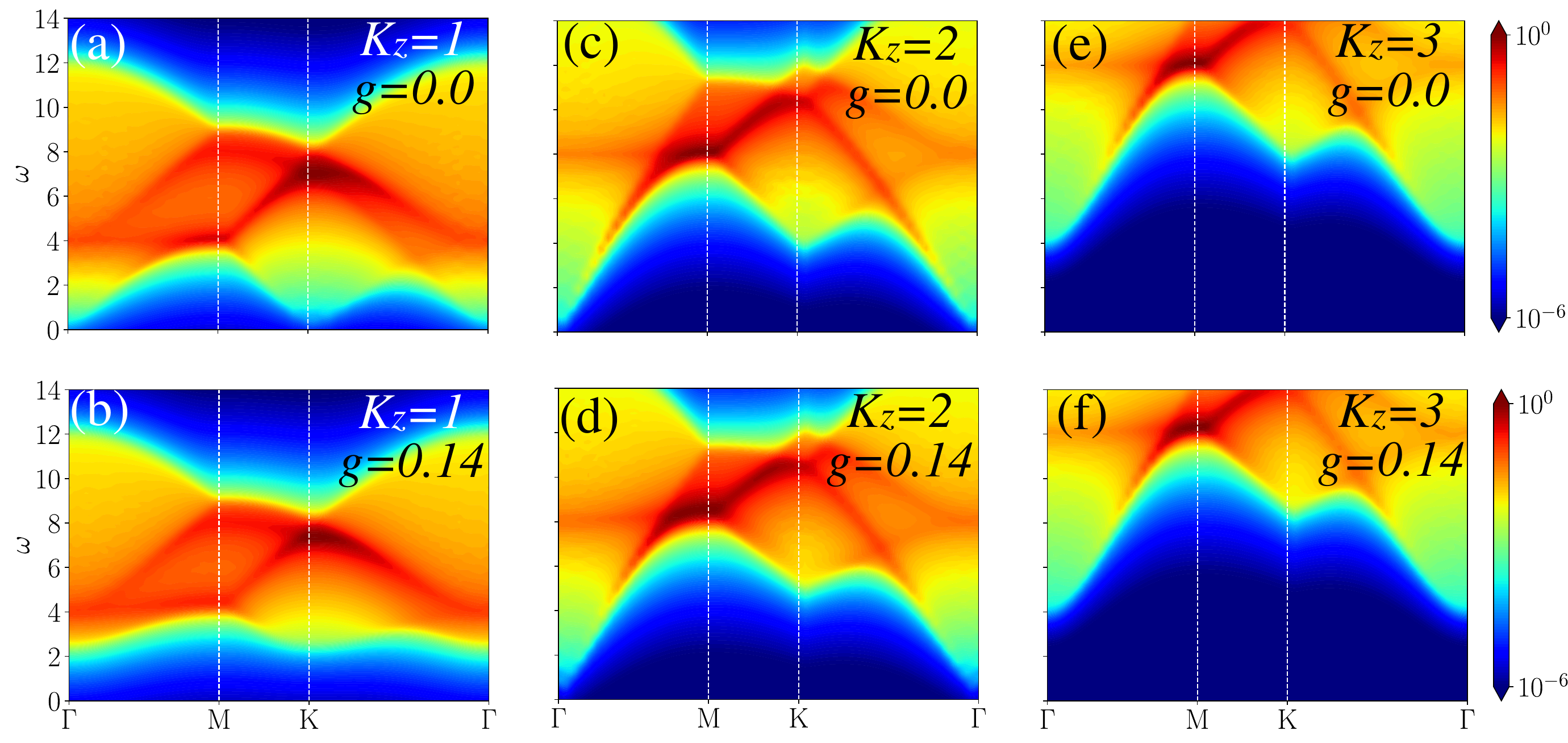}
\caption{ $S^{zz}_2(\mathbf{k},\omega)$ of two spin flip for AFM Kitaev model under varying magnetic field $h$ and different anisotropies. (a,b) Kitaev spin liquid at the isotropic limit, with and without the three-spin time-reversal-breaking perturbation $g$. (c,d) The same calculation done at the non-Abelian to Abelian transition $K_z/K = 2$, where $K \equiv K_x = K_y = 1$; and (e,f) inside the Abelian phase.  
\label{fig:twoflipapp}}
\end{figure*}
We used $a_{-\mathbf{q}}$ and  $b_\mathbf{q}$ are Majorana fermions modes of $A$ and $B$ sublattices.  Working in the zero-flux sector, it would be convenient to identify the effect of $\mathcal{D}_j^z \equiv \sigma_j^z \sigma_{j+z}^z$ on the pure Majorana sector. 
For clarity, we first apply Fourier transformation on $\mathcal{D}_j^z = \sigma_{i,A}^z \sigma_{i,B}^z$ without the parton decomposition, where we label $\sigma_j^z$ and $\sigma_{j+z}^z$ using their respective sublattice indices in the Bravais lattice, 
\begin{equation}
    \begin{split} \label{eq:ss}
    \mathcal{D}_\mathbf{k}^z &= {\rm F.T.}\{\sigma_{i,A}^z \sigma_{i,B}^z\} = \sum_i e^{-i \mathbf{k}\cdot \mathbf{r}_i} \sigma_{i,A}^z \sigma_{i,B}^z \\
    &    = \sum_{\mathbf{q}} \sigma_{\mathbf{k} - \mathbf{q}, A}^z \sigma_{\mathbf{q}, B}^z
    \end{split} 
\end{equation}
Furthermore, since $\mathcal{D}_\mathbf{k}^z$ does not affect the gauge sector, that is, as discussed in the main text 
\begin{align}
\mathcal{D}_j^z \ket{M_{0};\honey{\w}{\w}{\w}{\w}} &= i c_{j,A} c_{j,B} \ket{M_{0};\honey{\w}{\w}{\w}{\w}} \label{eq:dxflux1app}, 
\end{align}
where the $i$ attached to the Majorana bilinear comes from the definition in Eq.~\eqref{eq:bondf}. 
This leaves the gauge sector untouched, and its dynamical structure factor is completely determined by the Majorana sector. Therefore, we can define the $\tilde{\mathcal{D}}_\mathbf{k}^z$ operator that only describes the effect of $\mathcal{D}_\mathbf{k}^z$ on $M_0$ in a fixed uniform $Z_2$ gauge where all links are chosen to be $+1$. Hence, using the notation in Eq.~\eqref{eq:hamapp},  we have
\begin{equation} \label{eq:dk}
    \tilde{\mathcal{D}}_\mathbf{k}^z = {\rm F.T.}\{ic_j c_{j+z}\} = i\sum_\mathbf{q} a_{\mathbf{k} - \mathbf{q}} b_{\mathbf{q}}
\end{equation}
which is in keeping with Eq.~\eqref{eq:ss}.
In order to calculate the dynamical spectrum, we write it in terms of $C_{\mathbf{q},2}$, which according to Eq.~\eqref{eq:aqbq} takes the form:
\begin{equation}
\begin{split}
    \tilde{\mathcal{D}}_\mathbf{k}^z 
    = i \sum_\mathbf{q}F(\mathbf{k}-\mathbf{q}) \left(C_{{\mathbf{k}-\mathbf{q}},2} C_{\mathbf{q},2} + C_{-\mathbf{q},2}^\dagger C_{{-\mathbf{k}+\mathbf{q}},2}^\dagger \right)
\end{split}
\end{equation}
where we have defined
\begin{equation} \label{eq:tmpF}
    F(\mathbf{k}-\mathbf{q}) \equiv \frac{1}{2}\frac{\Delta_{\mathbf{k}-\mathbf{q}} - i\epsilon_{\mathbf{k}-\mathbf{q}}}{E_{\mathbf{k}-\mathbf{q}} - Q_{\mathbf{k}-\mathbf{q}}}
\end{equation}
and we have ignored terms like $C_{\mathbf{k}-\mathbf{q}}C_{-\mathbf{q}}^\dagger$ that do not contribute to dynamics (e.g. $C_{-\mathbf{q}}^\dagger\ket{\rm gs} = 0$). Hence, using Eq.~\eqref{eq:aqbq} and Eq.~\eqref{eq:dk}, the dynamical correlation becomes
\begin{widetext}
\begin{equation} \label{eq:nnkmapp}
\begin{split}
        S_{2}^{z}(k,\omega)
        &= \sum_{m \neq 0} \mel{0}{\mathcal{D}_\mathbf{k}^\alpha }{m}\mel{m}{\mathcal{D}_{-\mathbf{k}}^\alpha}{0} \delta(\omega - E_m + E_{\rm gs}) = -\sum_{m \neq 0} \mel{\rm gs}{\tilde{\mathcal{D}}^z_\mathbf{k}}{m} \mel{m}{\tilde{\mathcal{D}}^z_{-\mathbf{k}}}{\rm gs} \delta(\omega - E_m + E_{\rm gs}) \\
        &= -\sum_{m \neq 0} \sum_{\mathbf{q},\mathbf{p}} F(\mathbf{k}-\mathbf{q}) F(-\mathbf{k}-\mathbf{p})\bra{\rm gs} C_{{-\mathbf{q}},2}^\dagger C_{{-\mathbf{k}+\mathbf{q}},2}^\dagger \ket{m}\bra{m} C_{{-\mathbf{k}-\mathbf{p}},2} C_{\mathbf{p},2} \ket{\rm gs}\delta(\omega - E_m + E_{\rm gs})\\
        &= -\sum_{\mathbf{q}} F(\mathbf{k}-\mathbf{q}) F(-\mathbf{k}+\mathbf{q}) \delta[\omega -(E_{-\mathbf{k} + \mathbf{q}} + E_{-\mathbf{q}})] 
        \equiv \sum_{\mathbf{q}} G(\mathbf{k}-\mathbf{q}) \delta[\omega -(E_{-\mathbf{k} + \mathbf{q}} + E_{-\mathbf{q}})] 
\end{split}
\end{equation}
\end{widetext}
where $\ket{m} \equiv \ket{m(\mathbf{k}, \mathbf{q})}$ are excited states associated with quasi-particles with momentum $\mathbf{k}$ and $\mathbf{q}$; and we have defined $G$ using Eq.~\eqref{eq:tmpF}
\begin{equation} \label{eq:gk}
    G(\mathbf{k} - \mathbf{q}) \equiv -F(\mathbf{k} - \mathbf{q}) F(-\mathbf{k} + \mathbf{q}) = \frac{1}{4}\frac{E_{\mathbf{k}-\mathbf{q}}^2}{E_{\mathbf{k}-\mathbf{q}}^2 - Q_{\mathbf{k}-\mathbf{q}}^2}
\end{equation}
which is even under inversion as expected. Setting it in the infinite-lattice limit $N\rightarrow \infty$ and choosing the unit vectors of the lattice to be $\mathbf{n}_1 = (\frac{1}{2}, \frac{\sqrt{3}}{2}),\;\;
\mathbf{n}_2 =  (-\frac{1}{2},\frac{\sqrt{3}}{2})$, the continuous limit of Eq.~\eqref{eq:nnkmapp} thus takes on the form
\begin{equation} 
    S_2^{z} (\mathbf{k},\omega) = \frac{\sqrt{3}}{16 \pi^2} \int_{\rm BZ} G(\mathbf{k}-\mathbf{q}) \delta(\omega - \varepsilon_{\mathbf{k},\mathbf{q}}) d^2\mathbf{q}
\end{equation}
where $\varepsilon_{\mathbf{k},\mathbf{q}} \equiv E_{-\mathbf{k} + \mathbf{q}} + E_{-\mathbf{q}}$, 
as is used in the main text. 
At zero or infinitesimally small field, the distribution of $S_2^{z}(\mathbf{k}, \omega)$ is a direct consequence of two-particle density of states. Noting that, with zero or very small TR-breaking perturbation, $G(\mathbf{k})$ in Eq.~\eqref{eq:gk} would approximately become a constant $\simeq \frac{1}{4}$, and the distribution of $S_2^{z}$ in $\mathbf{k},\omega$ become determined only by the density of states of two-particle excitations. For example,  $S_2^{z}
(\mathbf{k} = \rm \Gamma, \omega \simeq 4)$ is a bright peak for $\mathbf{k} = 0$ cut, as shown in Fig.~\ref{fig:twoflipapp}. This is because, under zero or small perturbation, the largest two-particle density of states is at $\omega \simeq 2 E_{\mathbf{k} = 0}$, with $E_{\mathbf{k} = 0} \simeq 2$. 
Hence for pure Kitaev QSL or CSL with very small perturbation, we expect a bright spot at $\mathbf{k} = \Gamma,~ \omega \simeq 4$, as consistent in the iPEPS result in Fig.~2 of the main text. Similar argument can be used for other features. Importantly, as shown in Fig.~\ref{fig:twoflipapp}(b,c) the bright spot at ($\omega \simeq 7$, $ \mathbf{k} \simeq \rm K$) persists despite the perturbation, making it a sharp and robust feature that reflects the fractionalized quantum sector of Majorana fermions.

\begin{figure}[t] 
\centering
\includegraphics[width=0.45\textwidth]{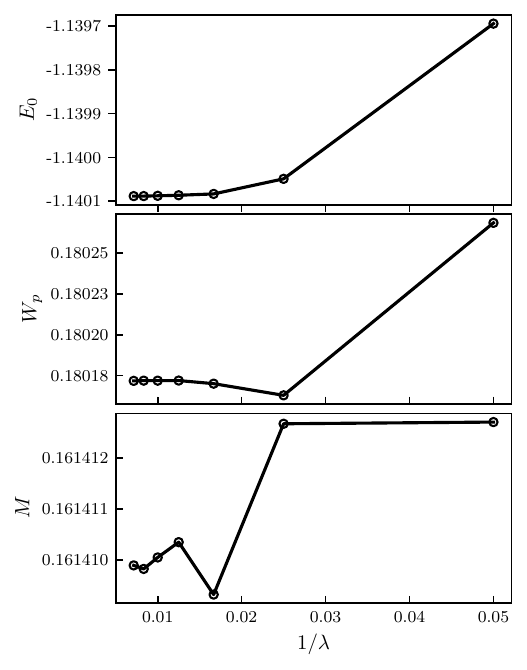}
\caption{The convergence of different observables with respect to boundary bond dimension $\lambda$.}
 
\label{fig:obsX}
\end{figure}
\begin{figure}[t] 
\centering
\includegraphics[width=0.45\textwidth]{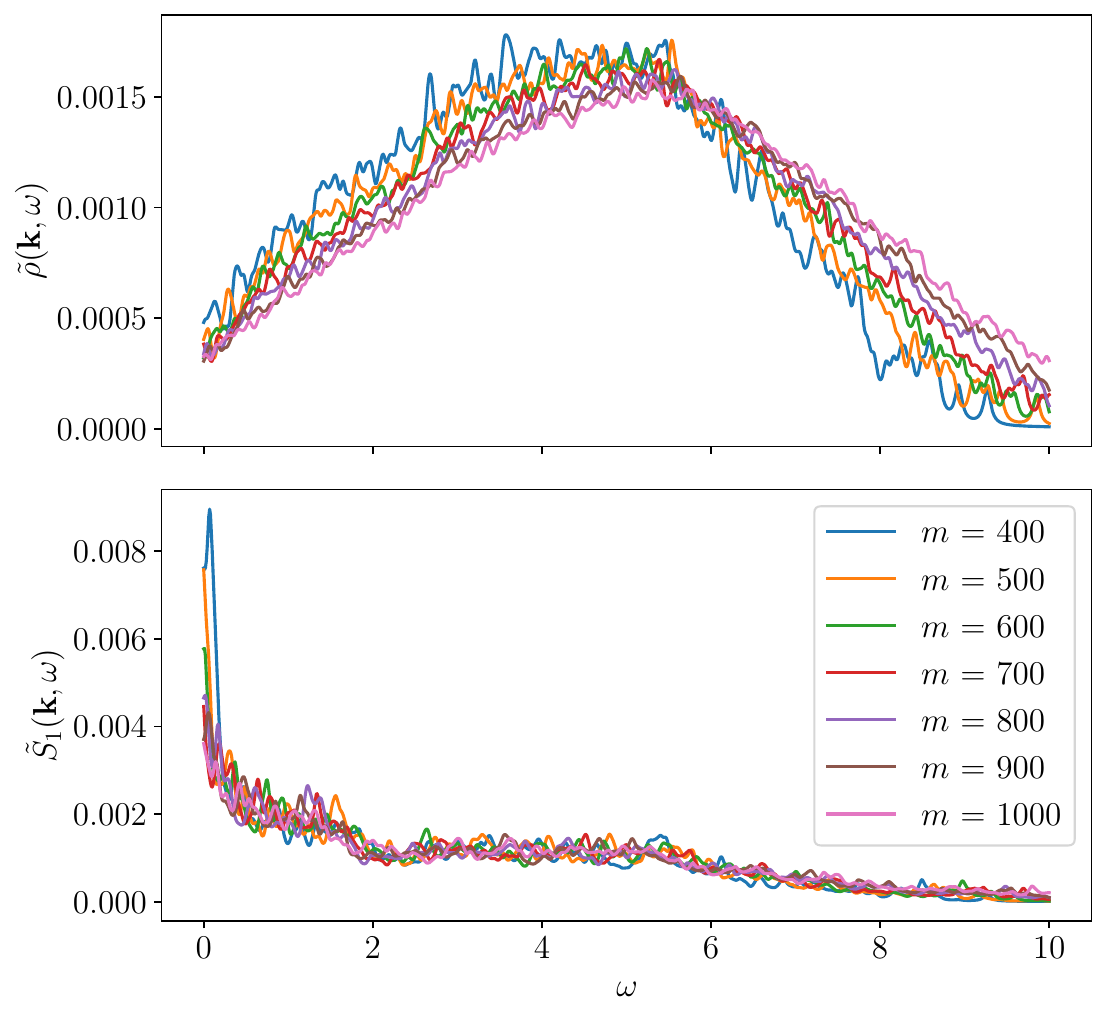}
\caption{The convergence of different observables with respect to boundary bond dimension $\lambda$.}
 
\label{fig:cut_h0.5M}
\end{figure}

Here in addition to the $S_{2}^{z}(\mathbf{k},\omega)$ presented in the main text at the isotropic limit of Kitaev couplings, we now list $S_{2}^{z}(\mathbf{k},\omega)$ for both the isotropic and anisotropic Kitaev coupling strengths, with and without the TR-breaking perturbation. Results are shown in Fig.~\ref{fig:twoflipapp}. These include both the non-Abelian (weak pairing) and the Abelian (strong pairing) phases as well as the topological phase transition at $K_z/K = 2,~K\equiv K_x = K_y = 1$. We note that the $S_2^{z}(\mathbf{k}, \omega)$ of the pure Kitaev QSL at the isotropic limit is the most susceptible to the TR-breaking perturbation. As shown in  Fig.~\ref{fig:twoflipapp}(a,b), the signal immediately above $\rm K$ point is abruptly pushed to higher energy upon introducing the next-nearest-neighbor hopping perturbation; while for the Abelian phase ($K_z/K > 2$) and the topological phase transition point ($K_z/K$ = 2), the next-nearest-neighbor hopping does not induce noticeable difference. This can be attributed to the fact that Majorana fermions are highly gapped in the anisotropic Abelian phase of the Kitaev honeycomb model; and the perturbation at the transition point does not gap out the Majorana fermion but simply alters the semi-Dirac point at $\rm M$ into a Dirac point at $\rm M$ \cite{Feng2023prb}, leaving no significant impact in energy levels of Majorana excitations. 

\begin{figure*}[t]
\centering
\includegraphics[width=0.95\linewidth]{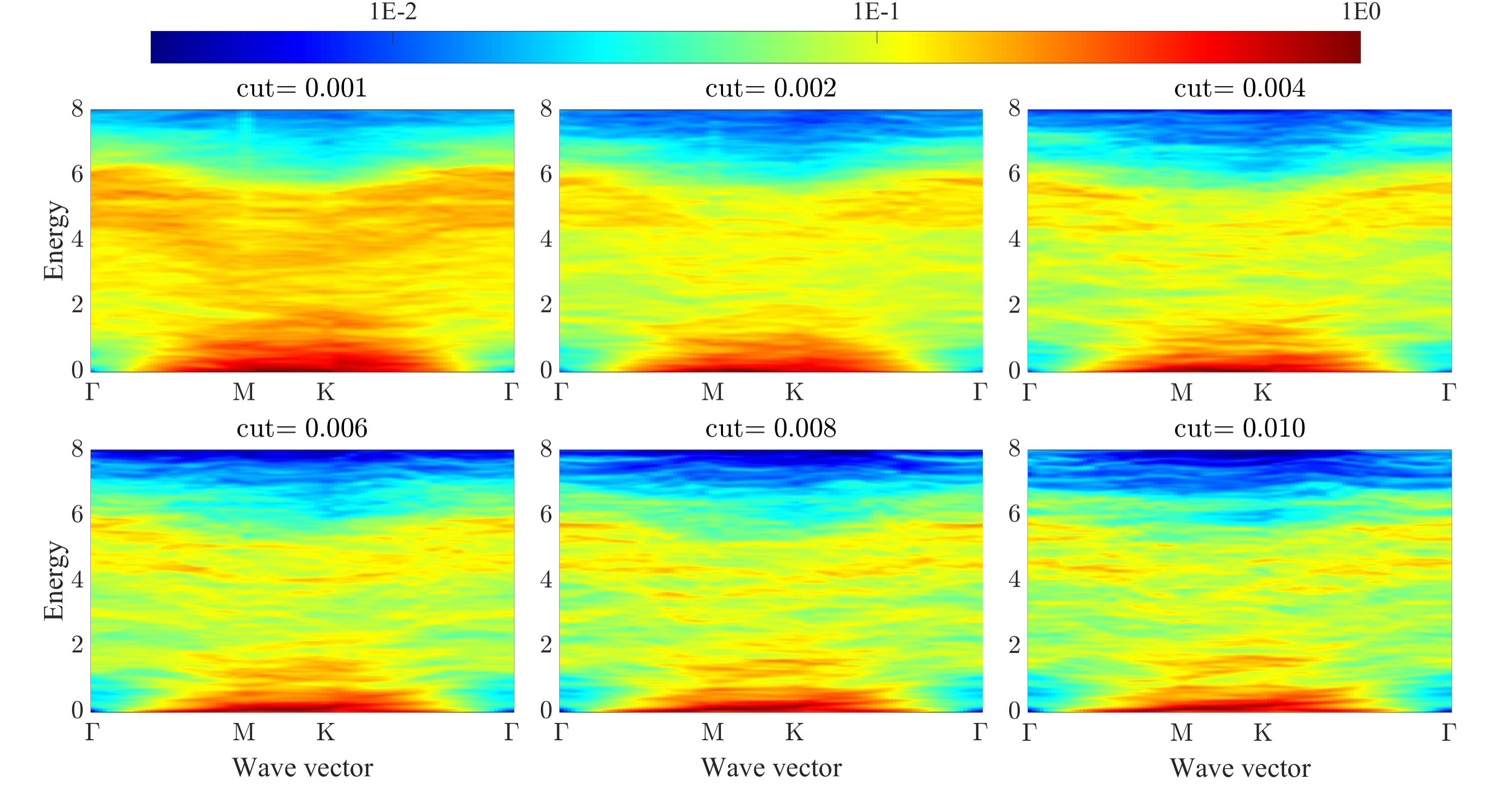}
\caption{Dynamical structure factor $S_1(\mathbf{k},\omega)$ for $h = 0.5$ with different cut thresholds. The corresponding number of basis varies from about 200 to 800.}
 
\label{fig:Spec_cut}
\end{figure*}
\begin{figure*}[t]
\centering
\includegraphics[width=0.95\linewidth]{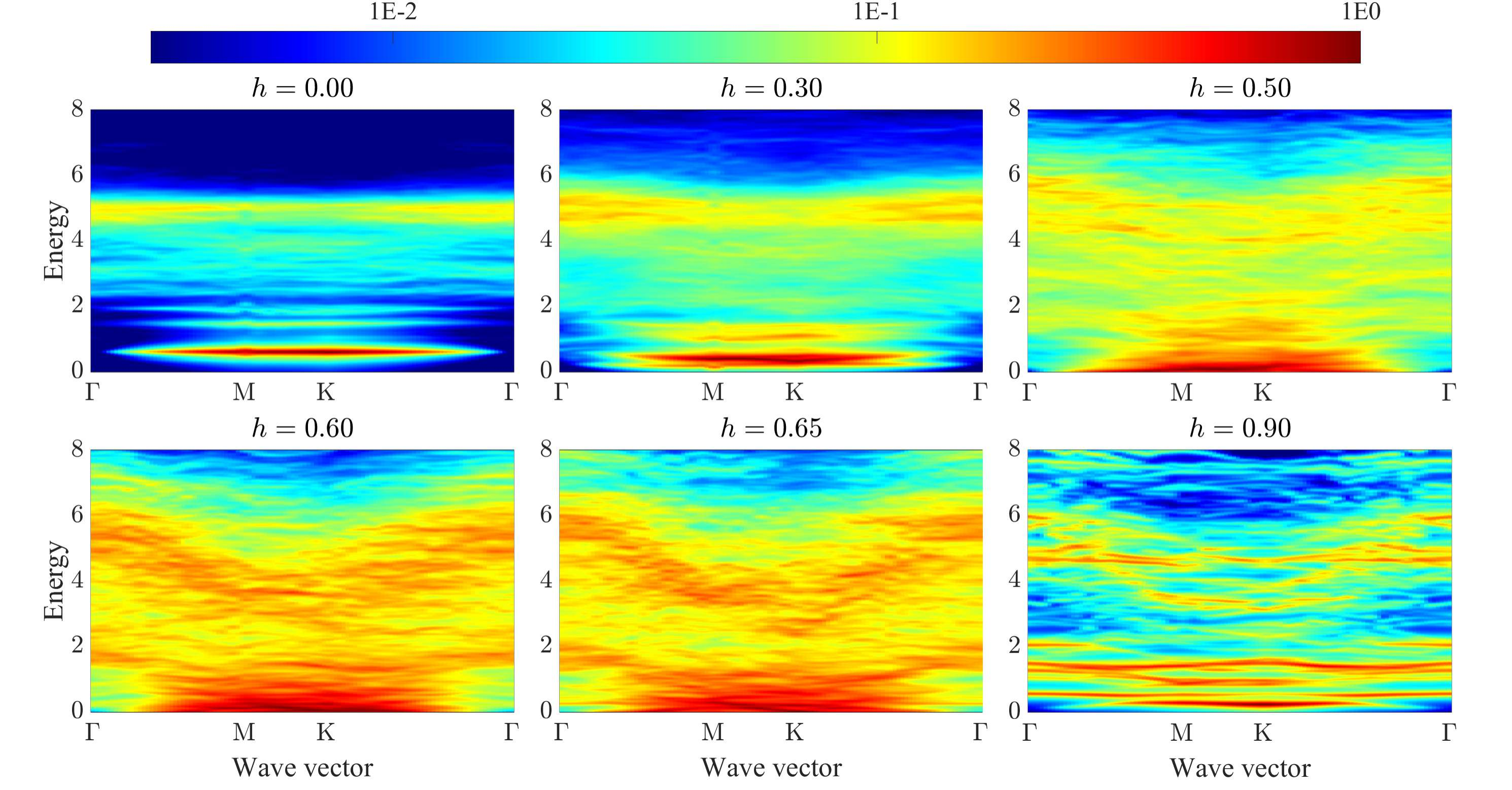}
\caption{Dynamical structure factor  $S_1(\mathbf{k},\omega)=\sum_{\alpha}S_1^{\alpha}(\mathbf{k},\omega)$ of one spin flip for AFM Kitaev model under varying magnetic field $h$. The spectra are presented on a logarithmic color scale along the momentum path $\rm\Gamma MK\Gamma$  through the BZ. 
}\label{oneflip}
\end{figure*}

\section{Convergence}

For completeness, we provide details about convergence in the numerical study in this section. In the CTMRG approach, ensuring convergence requires careful management of the bond dimension of both the local and boundary tensors, denoted as  $\lambda$ . Due to the high computational cost associated with spectrum calculations, we have limited the maximum bond dimension of the local tensor to  $D=5$. In Fig.~\ref{fig:obsX}, we demonstrate the convergence of the ground state observables with respect to  $\lambda$.

\begin{figure*}[t]
\centering
\includegraphics[width=0.95\linewidth]{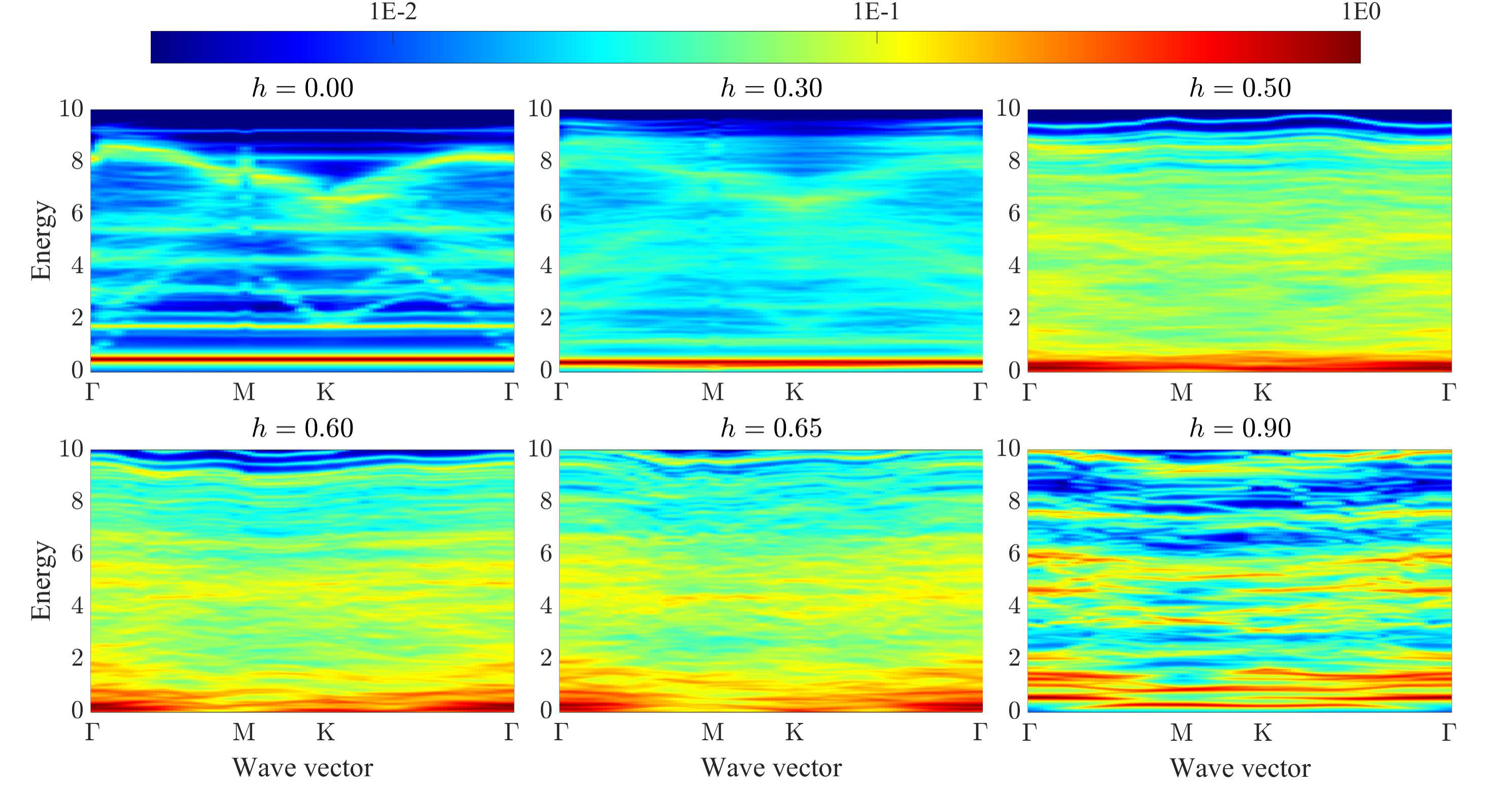}
\caption{Dynamical structure factor  $S_2(\mathbf{k},\omega)=\sum_{\alpha}S_2^{\alpha}(\mathbf{k},\omega)$ of two spin flip for AFM Kitaev model under varying magnetic field $h$. The spectra are presented on a logarithmic color scale along the momentum path $\rm\Gamma MK\Gamma$  through the BZ. 
}\label{twoflip}
\end{figure*}
\begin{figure*}[t]
\centering
\includegraphics[width=0.85\linewidth]{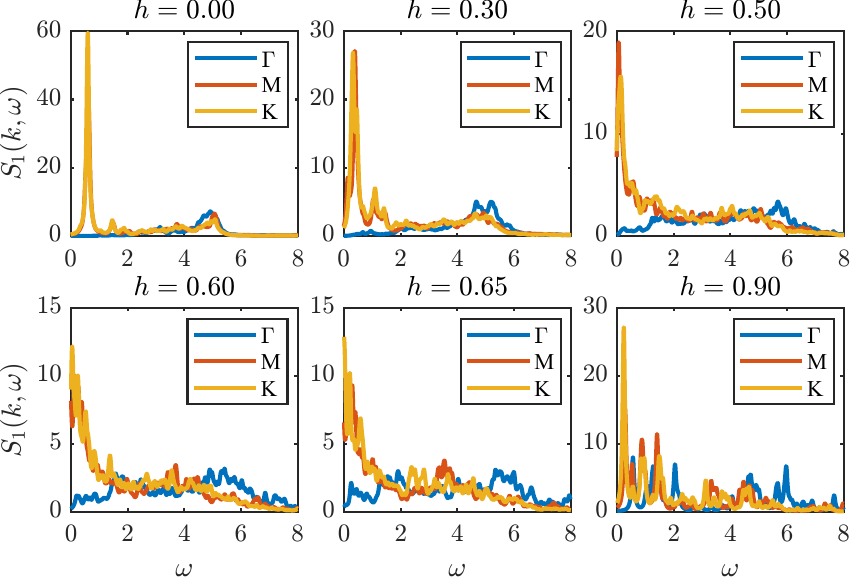}
\caption{$S_1(\mathbf{k},\omega)$ at $\rm\Gamma, M, K$ points. The broadening factor is $\eta=0.05$.
}\label{S1cuts}
\end{figure*}
\begin{figure*}[t]
\centering
\includegraphics[width=0.85\linewidth]{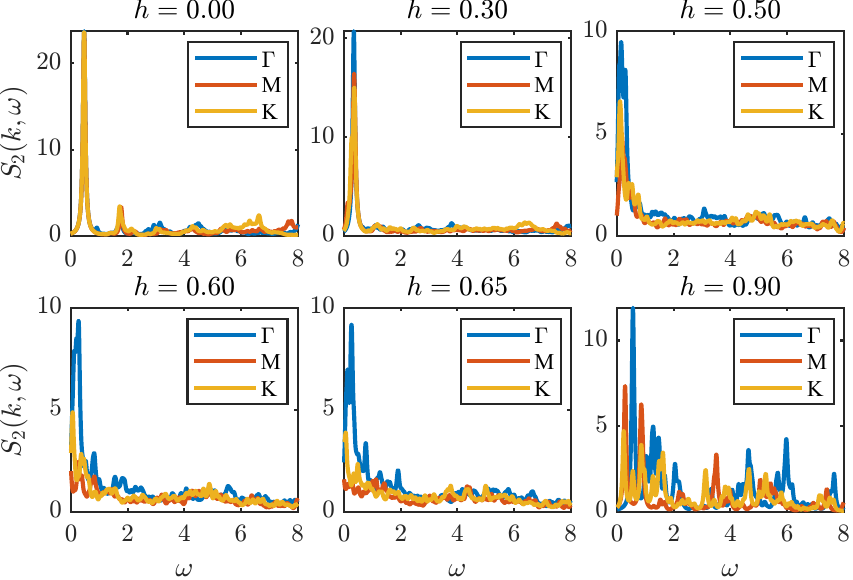}
\caption{$S_2(\mathbf{k},\omega)$ at $\rm\Gamma, M, K$ points. The broadening factor is $\eta=0.05$.
}\label{S2cuts}
\end{figure*}
Additionally, the excitations can become unstable if the norm matrix is ill-conditioned. We therefore reduce the number of basis functions when evaluating the effective Hamiltonian. This is done by diagonalizing the norm matrix and approximating it as 
\begin{equation}
    N = P^{-1}\Lambda P\approx \tilde{P}^{-1}\tilde{\Lambda}\tilde{P},
\end{equation}
where $\tilde{\Lambda}$, 
and $\tilde{P}$ are constructed from $m$ largest eigenvalues.  The projected effective Hamiltonian is thereafter defined as
\begin{equation}
    \tilde{H} = \tilde{P}H\tilde{P}^{-1}
\end{equation}
Larger $m$ will lead to more energy levels and higher accuracy. However, some spurious eigenlevels will appear due to the instability. Here, we present the normalized DOS $\tilde{\rho}_1(\mathbf{k},\omega)$ and $\tilde{S}_1(\mathbf{k},\omega)$ respect to $m$ normalized using a single integral over $\omega$ to show the convergence and stability. For instance, we show results for $M$ point at $h=0.5$ in Fig.~\ref{fig:cut_h0.5M}. Notably, we do not fix $m$ for different momenta; instead, we set a threshold for the eigenvalues, leading to slight fluctuations in $m$ for different $\mathbf{k}$.  To demonstrate this point, we show the final spectra results in Fig.~\ref{fig:Spec_cut} for various thresholds, where the results remain robust under the change of the threshold. The results in the main text are all set with threshold as $0.006$, with about $320$ basis retained.

\begin{figure*}[t]
\centering
\includegraphics[width=\linewidth]{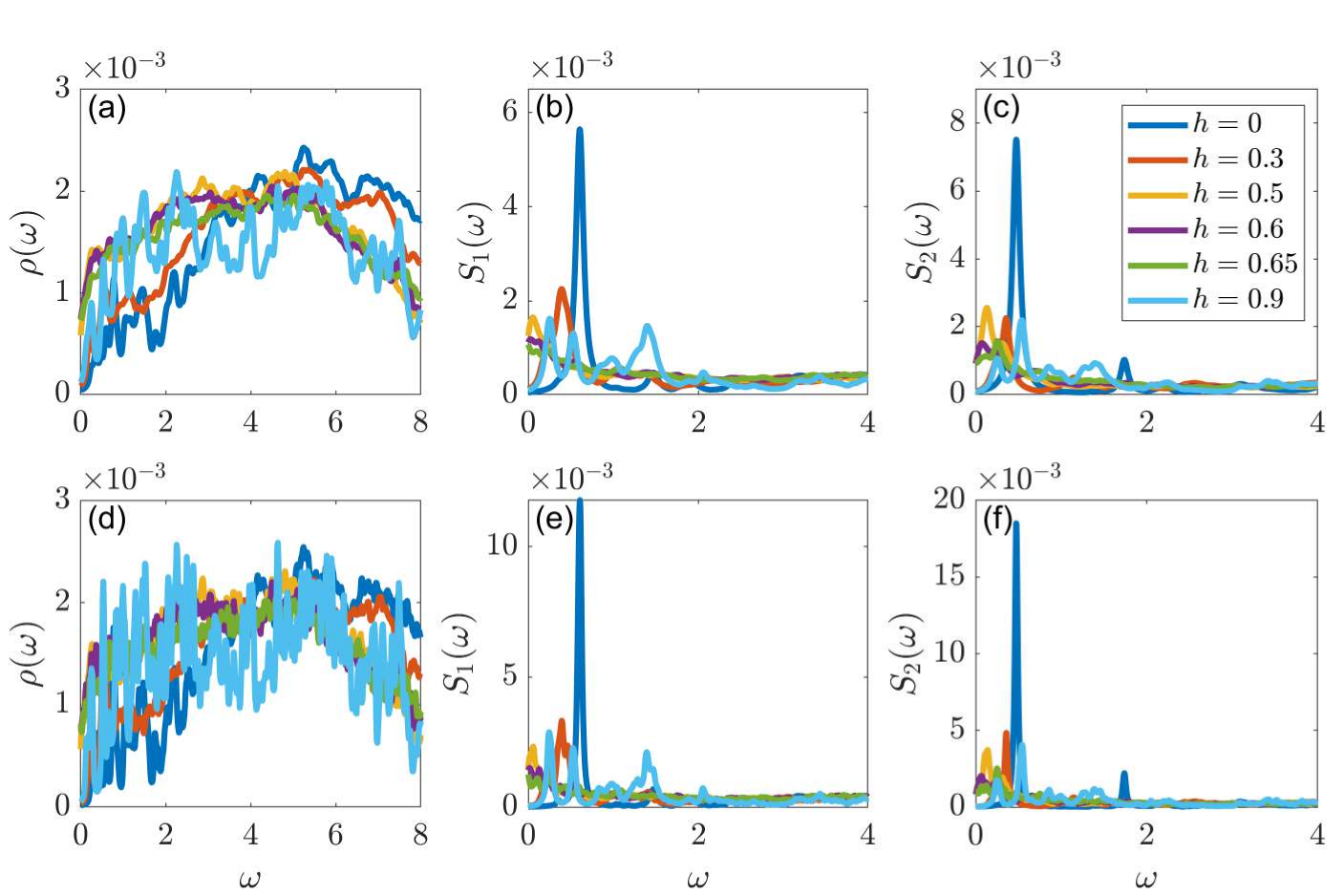}
\caption{ iPEPS results for (a)(d) the density of states $\rho(\omega)$, (b)(e) the intergrated single spin flip spectra $S_1(\omega)$, and (c)(f) the intergrated single spin flip spectra $S_2(\omega)$.
The broadening factor is $\eta = 0.05$ for the upper three figures and $\eta = 0.02$ for the lower three figures. Notably, at $\eta = 0.02$ (as well as $\eta = 0.01$ shown in Fig.~4(c) in the main text),  we still find no observable gap nor the trend of opening a gap in $S_1(\omega)$, $S_2(\omega)$ and $\rho(\omega)$ at the lowest energies which are comparable to or lower than previous putative gaps, ranging from $0.01$ to $0.05$, obtained by parton mean field theories \cite{Jiang2020,ZhangNatComm2022}. This suggests that although quadratic parton mean-field theories are capable of qualitatively capturing phase transitions, the apparent gap in the non-perturbative regime can be spurious due to the missing many-body entanglement between bond fermions, which nevertheless remains intact in iPEPS calculations. 
}
\label{fig:DOS}
\end{figure*}

\section{Additional Results}
Here we provide the spectra for additional parameters, as shown in Fig.~\ref{oneflip} for the one-spin flip spectra $S_1(\mathbf{k},\omega)$ and Fig.~\ref{twoflip} for the two-spin flip spectra $S_2(\mathbf{k},\omega)$. The spectra at the $\Gamma$, $\rm M$, and $\rm K$ points are shown in Fig.~\ref{S1cuts} and Fig.~\ref{S2cuts}.
To demonstrate the intermediate phase is gapless, we calculate the approximate density of states (DOS) and integrated spectra along the $\Gamma\rm M K\Gamma$ path, defined as follows:
\begin{align}
    &\rho(\omega) = \frac{\int_{\rm \Gamma MK\Gamma}d\mathbf{k}\sum_m\delta(\omega-E_\mathbf{k}^m+E_{\rm gs})}{\int_0^\infty d\omega\int_{\rm\Gamma MK\Gamma}d\mathbf{k}\sum_m\delta(\omega-E_\mathbf{k}^m+E_{\rm gs})}\\
    &S_1(\omega) = \frac{\int
_{\rm\Gamma MK\Gamma}d\mathbf{k}S_1(\mathbf{k,\omega})}{\int_0^\infty d\omega\int_{\rm\Gamma MK\Gamma}d\mathbf{k}S_1(\mathbf{k,\omega})}\\
    &S_2(\omega) = \frac{\int_{\rm\Gamma MK\Gamma}d\mathbf{k}S_2(\mathbf{k,\omega})}{\int_0^\infty d\omega\int_{\rm\Gamma MK\Gamma}d\mathbf{k}S_2(\mathbf{k,\omega})}.
\end{align}
The results are shown in Fig.~\ref{fig:DOS}. 

\clearpage


\end{widetext}

\end{document}

%% file: header.tex
\AtBeginDocument{
\heavyrulewidth=.08em
\lightrulewidth=.05em
\cmidrulewidth=.03em
\belowrulesep=.65ex
\belowbottomsep=0pt
\aboverulesep=.4ex
\abovetopsep=0pt
\cmidrulesep=\doublerulesep
\cmidrulekern=.5em
\defaultaddspace=.5em
}

\usepackage{float}
\usepackage{comment}
\usepackage{booktabs}
\usepackage[percent]{overpic}
\usepackage{microtype}
\usepackage{array}
\usepackage{physics}
\usepackage{graphicx}
\usepackage{bbold}
\usepackage[usenames,dvipsnames]{xcolor}
\usepackage{tikz}
\usepackage{epsfig}
\usetikzlibrary{arrows.meta, automata, positioning, quotes, shapes}

\bibpunct{[}{]}{,}{n}{}{}

\newcommand{\honey}[4]{
	\def \w {white}
	\def \g {lightgray}
	\begin{tikzpicture}[baseline={([yshift=-.5ex]current bounding box.center)}, scale=0.5]
	  \begin{scope}[%
	every node/.style={anchor=west,
	regular polygon, 
	regular polygon sides=6,
	draw,
	minimum width=1cm,
	outer sep=0,
	},
	      transform shape]
	    \node (A)[fill=#1] 	    	        {};
	    \node (B)[fill=#2]  at (A.corner 5) {};
	    \node (C)[fill=#3]  at (A.corner 1) {};
	    \node (D)[fill=#4]  at (C.corner 5) {};
	  \end{scope}
	\end{tikzpicture}
}

\definecolor{ForestGreen}{HTML}{668000}
\definecolor{red1}{HTML}{FF4136}
\definecolor{green1}{HTML}{00802b}


%% file: main.bbl
\begin{thebibliography}{105}
\bibitem [{foo()}]{footnote}%
  \BibitemOpen
  \href@noop {} {\bibinfo {title} {Our calculation suggests that the {CSL} to
  {IGP} transition is continuous. {Further} finite-size scaling analysis is
  needed to firmly establish this claim, which is beyond the scope of this
  work.}}\BibitemShut {Stop}%
\end{thebibliography}
